\documentclass[aip,jcp, numerical, preprint]{revtex4-1}
\usepackage{graphicx}
\graphicspath{{images/}}
\usepackage[table,xcdraw]{xcolor}
\usepackage{amsmath,amsfonts}
\usepackage[cal=cm, scr=rsfso]{mathalfa}
\usepackage{newtxtext,newtxmath,bm}
\usepackage[version=3]{mhchem}
\usepackage{float}
\usepackage{enumitem}
\usepackage{booktabs}
\usepackage{makecell}
\usepackage{array} 
\usepackage{tabularx}

\usepackage{pifont}
\newcommand{\cmark}{\ding{51}}%
\newcommand{\xmark}{\ding{55}}%

\begin{document}
\newcommand \e[1]{{\textbf{#1}}}
\newcommand \bo[1]{{\bf{{#1}}}}
\newcommand \bra[1] {\left\langle {#1}  \right\vert}
\newcommand \ket[1] {\left\vert {#1} \right\rangle }
\newcommand \braket[2] { \left\langle {#1}  \middle\vert{#2} \right\rangle}
\newcommand \ketbra[2] { \left\vert {#1}  \middle\rangle  \middle\langle {#2} \right\vert}
\newcommand \braketthree[3] { \left\langle {#1}  \middle\vert{#2} \middle\vert{#3} \right\rangle}
\newcommand{\RNum}[1]{\uppercase\expandafter{\romannumeral #1\relax}}


\title{On the Electronic Path-Integral Normal Modes of the Spin-Mapping Representation of Nonadiabatic Dynamics}
\author{Lauren E. Cook}
\author{James R. Rampton}
\author{Timothy J. H. Hele}
 \email{\href{mailto:t.hele@ucl.ac.uk}{t.hele@ucl.ac.uk}}
 \affiliation{Department of Chemistry, University College London, Christopher Ingold Building, London WC1H 0AJ, United Kingdom}

\date{\today}

\begin{abstract}
Nonadiabatic Matsubara dynamics has been proposed in both the Meyer--Miller--Stock--Thoss (MMST) and spin-mapping representations, whereby the dynamics are truncated in the higher path-integral normal modes of the nuclear degrees of freedom, but no truncation is performed in the electronic variables. In contrast to single-surface Matsubara dynamics, these methods do not appear to conserve the Quantum Boltzmann Distribution (QBD) for general systems. Recently, it was shown that truncating in the electronic path-integral normal modes of the MMST representation does not lead to conservation of the QBD for a single trajectory or accurate computation of a correlation function. Here, the electronic path-integral normal modes in the \emph{spin-mapping} representation are investigated. We focus on a two-level system although we expect our findings to be applicable to any number of electronic states. We find, perhaps surprisingly, that the higher normal modes of the spin-vector $z$-component, the difference in state populations, are constrained by the QBD, that the time-evolved state population observable is a function of only the spin-mapping centroid, and that the full $N$-bead computational results are replicated using only the spin-mapping centroid, leading to accurate, QBD conserving dynamics. While at present these results are not a generally-applicable method, we believe this may explain the recent success of spin-mapping approaches compared to MMST approaches, 
and should aid the future derivation of highly accurate and Quantum Boltzmann conserving nonadiabatic dynamics methods.

\end{abstract}


\maketitle 

\allowdisplaybreaks

\section{Introduction}
Simulation of nonadiabatic dynamics, crucial for understanding light and energy transfer, is particularly challenging to implement due to the coupled nuclear and electronic degrees of freedom (dof). Full quantum methods, such as Multi-Configurational Time-Dependent Hartree Fock (MCTDH) and wavepacket methods,\cite{Beck2000,vanhaeftenPropagatingMultidimensionalDensity2023,wangMultilayerFormulationMulticonfiguration2003, shinMultipleTimeScale1996, vanhaeftenPropagatingMultidimensionalDensity2023, mukherjeeAssessingNonadiabaticDynamics2025} are extremely accurate but limited to smaller systems by the large associated computational expense. For single surface (adiabatic) systems, many classical-scaling methods have been introduced including ring-polymer molecular dynamics (RPMD),\cite{Craig2004, Habershon2013, Hele2015, heleAlternativeDerivationRingpolymer2016} centroid molecular dynamics (CMD),\cite{ Cao1993, Cao1994, Cao1994a, Cao1994b,heleCommunicationRelationCentroid2015, jangPathIntegralCentroid1999, jangDerivationCentroidMolecular1999, Jung2020, castro_vibrational_2025} and thermostatted (T)-RPMD,\cite{Rossi2014, Hele2015c, Hele2016a,hele_thermal_2017, ceriottiEfficientStochasticThermostatting2010} and the Linearised Semiclassical Initial-Value Representations (LSC-IVR).\cite{ Stock1997, Miller1970,Sun1998, Sun1998b, Wang1998, Wang1999} Several extensions for nonadiabatic dynamics are available such as mean-field RPMD, nonadiabatic (N)-RPMD and mapping variable (MV)-RPMD for RPMD,\cite{Hele2011, Richardson2013, Richardson2017, Ananth2013, pierre_mapping_2017} and Mixed Quantum Classical (MCQ)-IVR for IVR methods.\cite{Ananth2007, Church2017,Church2018, Antipov2015, Liu2015, Filinov1986, Thoss2001, Shi2004} Additionally, several mixed quantum-classical methods exist including Ehrenfest methods, surface hopping and mapping methods.\cite{Kapral2016, Tully1971, tullyPerspectiveNonadiabaticDynamics2012, Shakib2017, Shalashilin2011, Zimmermann2014, Meyer1979,Stock1997, Stock2005, Runeson2019, MASH, althorpeNonadiabaticReactionsGeneral2017, Althorpe2016,nelsonNonadiabaticExcitedStateMolecular2020} However, all these methods have their limitations and in our view no single method has emerged as the optimal choice for simulating nonadiabatic dynamics. 

One particularly desirable method property for a nonadiabatic dynamics method is conservation of the quantum Boltzmann distribution (QBD) in order to avoid the system drifting from thermal equilibrium which, for example, could result in incorrect physical observables and artificial energy transfer at long times \cite{habershon_zero_2009}. For single surface systems, Matsubara dynamics conserves the QBD through truncation in the higher nuclear imaginary-time path-integral normal modes.\cite{Willatt2017, Hele2015} Only the lowest (smooth) Matsubara modes are propagated whereas the dynamics of the higher (non-smooth) normal modes is discarded.\cite{Hele2013, treninsMeanfieldMatsubaraDynamics2018, Hele2015} For operators corresponding to physical observables, the observable is usually a function of a finite number of the \emph{lowest} normal modes (as has previously been noted for the dividing surface in instanton theory and quantum transition-state theory).\cite{Hele2013,Althorpe2013,Hele2013b,heleAlternativeDerivationRingpolymer2016, Hele2014,Richardson2009} Under the Matsubara dynamics approximation, the dynamics of the smooth modes becomes classical, conserves the QBD and is generally a very good approximation to the exact quantum time-correlation function.\cite{Hele2015,heleCommunicationRelationCentroid2015}   

For multiple surfaces, Rabi oscillations can occur between electronic states that contribute to nonadiabatic effects such that it is desirable to obtain these simultaneously with QBD conservation.\cite{cookElectronicPathIntegral2025} One way to obtain Rabi oscillations is to propagate using a mapping Hamiltonian, which is the class of methods focused on in this work.\cite{Meyer1979, Stock1997, Runeson2019} These trajectory-based methods map the discrete quantum system onto continuous variables which are propagated classically. To investigate dynamical properties of nonadiabatic systems, mapping methods are often utilised to approximate correlation functions of the form 
\begin{align}
    \label{General-CF}
    C_{AB}= \frac{1}{Z} \mathrm{Tr}[e^{-\beta \hat{H}}\hat{A}(0)\hat{B}(t)]\text{,}
\end{align}
where $e^{-\beta \hat{H}}$ is the quantum Boltzmann distribution at inverse temperature, $\beta = 1/ k_B T$, $\hat{A}$ and $\hat{B}$ are operators and $Z$ is the partition function given by
\begin{align}
    \label{partiton}
    Z = \mathrm{Tr} [e^{-\beta \hat{H}}] \text{.}
\end{align}

One well-established mapping is the Meyer-Miller-Stock-Thoss (MMST) mapping which obtains electronic positions and momenta and evolves under the MMST Hamiltonian using Hamilton's equations of motion (EOM).\cite{Meyer1979, Stock1997, Church2018, Cook2023} Alternatively, spin-mapping obtains a spin-vector on the surface of a Bloch sphere (in the two-level case) which follows Heisenberg's EOM.\cite{Runeson2019, Runeson2020, Runeson2021} This is advantageous as it removes two redundant dof compared to the MMST representation.\cite{Runeson2019, Runeson2020} Spin-mapping is also invariant to any choice of Hamiltonian splitting, which is a known issue for the MMST mapping.\cite{Bossion2021, Richardson2017, Runeson2019} Originally, spin-mapping was limited to two electronic states,\cite{Runeson2019, Bossion2021} but spin-coherent states and the $SU(F)$ Lie group have been harnessed to generalise to $F>2$ states.\cite{Runeson2020, bossionNonadiabaticMappingDynamics2022, bossionNonadiabaticRingPolymer2023} The MMST and spin-mapping representations are related by a transformation and the Hamiltonians are equivalent under the addition of a zero-point energy (ZPE) parameter to the MMST Hamiltonian.\cite{Runeson2019, Runeson2020, bossionNonadiabaticMappingDynamics2022} ZPE fitting parameters were first introduced by Stock and M\"uller to mitigate ZPE leakage, when energy unphysically flows between states.\cite{Muller1999, Stock1999} However, spin-mapping derives this as a function of the number of states and leads to a parameter that is close to what was previously found to be optimal.\cite{Runeson2019}  

Due to the apparent high accuracy of spin-mapping approaches compared to MMST, spin-mapping has gained a significant amount of interest in recent years. This resulted in the development of many successful spin-mapping methods including a partially linearized approach,\cite{Mannouch2020, Mannouch2020b, Mannouch2022} an ellipsoidal method,\cite{Amati2023} generalisation to multiple states,\cite{bossionNonadiabaticMappingDynamics2022, bossionNonadiabaticRingPolymer2023, Runeson2020} the Spin-MInt algorithm,\cite{cookSpinMIntAlgorithmAccurate2026,rampton_symplectic_2026} the Mapping Approach to Surface Hopping (MASH),\cite{MASH, lawrenceSizeconsistentMultistateMapping2024, geutherTimeReversibleImplementationMASH2025,richardsonNonadiabaticDynamicsMapping2025, runeson_exciton_2024, furlanettoSimulatingElectronicCoherences2025} and spin-mapping (SM)-NRPMD,\cite{Bossion2021, bossionNonadiabaticRingPolymer2023}. While we have detailed some of the known advantages of spin-mapping over MMST here, we believe that the higher accuracy of spin-mapping has not yet been fully explained and deserves further investigation.


For nonadiabatic mapping methods, it is particularly challenging to obtain conservation of the QBD and Rabi oscillations simultaneously.\cite{cookElectronicPathIntegral2025, Ananth2013, Richardson2013} While the ellipsoidal spin-mapping replicates Rabi oscillations and preserves the QBD, its mean-field dynamics was found to be often less accurate for short times than the original spin-mapping method and in its current formalism, can only be utilised for a two-level system.\cite{Amati2023} Nonadiabatic Matsubara dynamics has been proposed in both the MMST representation (NA-Mats) and the spin-mapping representation (SM-NA-Mats) but, to our knowledge, neither method conserves the QBD exactly.\cite{Chowdhury2021, Richardson2017, bossionNonadiabaticRingPolymer2023} This is surprising as these approaches follow the Matsubara approximation and truncate in the nuclear path-integral normal modes although the electronic dof are left unchanged. One can therefore pose the question of whether additionally following the Matsubara approach with electronic path-integral normal modes will lead to QBD conservation for a nonadiabatic Matsubara method. A recent study of the electronic path-integral normal modes of the MMST representation showed that such a method is unlikely to be obtained through use of MMST variables and the related electronic state descriptors.\cite{cookElectronicPathIntegral2025} This is as truncation in the higher electronic path-integral normal modes of the MMST representation does not result in QBD conservation or accurate dynamics and there is no constraint by the QBD on the higher MMST normal mode distributions.\cite{cookElectronicPathIntegral2025} In addition, the physical observable (such as electronic state) is generally a function of \emph{all} MMST normal modes rather than a finite number of the lowest modes.\cite{cookElectronicPathIntegral2025} However, this does not mean that all choices of electronic metrics are unsatisfactory and several alternative metrics could be investigated, such as spin-mapping and action-angle variables.\cite{cookElectronicPathIntegral2025}

In this article, we investigate the electronic path-integral normal modes of the spin-mapping representation. For simplicity, we focus on an electronic-only system and rather surprisingly find that the observable is a function of only the lowest normal mode, the higher spin-mapping normal modes are constrained by the QBD and accurate, QBD conserving dynamics is obtained using \textit{only} the lowest spin-mapping normal mode. These properties were not obtained for the MMST normal modes.\cite{cookElectronicPathIntegral2025} While these results are not yet a widely-applicable method, they may provide reasoning behind the recent success of spin-mapping compared to MMST and upon reintroduction of nuclear variables and in future research could result in the first QBD conserving nonadiabatic Matsubara method. 

The structure is as follows. In Section \ref{backgroundtheory}, we present the background theory on spin-mapping and Matsubara dynamics. In Section~\ref{methodology}, we determine that the metric constrained by the QBD is the electronic normal modes of the difference in state populations, which is naturally encoded in spin-mapping. Hence, we present the spin-mapping electronic normal modes, finding that the electronic-only observables becomes a function of only the lowest normal mode, related to the centroid, and this truncation conserves the QBD. In Section~\ref{chap3-results}, we computationally confirm these algebraic results. We conclude in Section~\ref{conclusions}.

\section{Background Theory} \label{backgroundtheory}
\subsection{Spin-Mapping} \label{spin-mapping-BT}
The general quantum Hamiltonian operator for multiple electronic potential energy surfaces is
\begin{align} \label{quantum-ham}
    \hat{H} = \hat{T}(\hat{\bo{P}})+ \hat{{V}}(\hat{\bo{R}}) \textrm{,}
\end{align}
where $\hat{T}$ and $\hat{V}$ are the kinetic and potential energy operators respectively and, $\hat{\bo{R}}$ and $\hat{\bo{P}}$ are the nuclear position and momentum operators. Key electronic configurations for photochemistry are avoided crossings (where states become close in energy but
do not cross) and conical intersections (where states are degenerate in energy).\cite{patersonConicalIntersectionsPerspective2005, worthBORNOPPENHEIMERMolecularDynamics2004} By transforming to the diabatic representation, one can propagate through these regions to observe nonadiabatic effects, such that for an \textit{F}-level electronic system\cite{Stock1997, Meyer1979} 
\begin{equation} \label{ODH}
	\hat{H}= \frac{1}{2}\hat{\bo{P}}^\mathrm{T} \boldsymbol{\mu}^{-1}\hat{\bo{P}} + \sum_{n,m=1}^{F} \ket{n} \bar{\bo{V}}_{nm}(\hat{\bo{R}}) \bra{m} \text{,}
\end{equation}
where $\bar{\bo{V}}(\hat{\bo{R}})$ is an $F \times F$ diabatic electronic potential energy matrix in the basis of the electronic states, $\{\ket{n}\}$, and $\boldsymbol{\mu}$ is a $K \times K$ diagonal matrix of nuclear masses. Throughout this work, we set $\hbar =1$.
The diabatic potential, $\bar{\bo{V}}$, can be split into state-independent and state-dependent potentials, $U(\bo{R})$ and $\bo{V}(\bo{R})$.\footnote{This often improves the stability of propagation under the MMST Hamiltonian}\cite{Kelly2012} For a two-level system, $\bo{V}(\bo{R})$ typically has the following form 
\begin{align} \label{diabatic pot mat}
    \bo{V}(\bo{R}) = \begin{bmatrix}
        V_1(\bo{R}) & \Delta^*(\bo{R}) \\
        \Delta(\bo{R}) & V_2(\bo{R})
    \end{bmatrix} \text{,}
\end{align}
where $\Delta$ is the coupling between electronic states, $^*$ represents the complex conjugate and $V_{1/2}$ are the state energies.

\begin{figure}[t]
        \centering
        \includegraphics[width=0.5\linewidth]{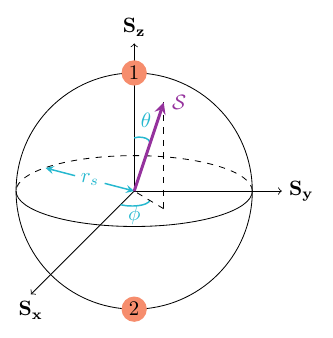}
        \caption[Bloch sphere diagram of spin-vector.]{The expectation values of the spin-operators form a basis of Cartesian coordinates where a Bloch sphere of radius, $r_s = 1/2$, can be defined. The motion of the spin-vector (purple) on the surface is circular around the Hamiltonian vector. When the spin-vector is on either of the poles (orange circle), the population is entirely in one state. }
        \label{fig:spin-mapping}
\end{figure}

Spin-mapping maps a two-level quantum system onto a classical spin-vector on a Bloch sphere by using the equivalence to a spin-1/2 particle in an external magnetic field (which is in a one-to-one correspondence with the Hamiltonian), as seen in Figure~\ref{fig:spin-mapping} if the radius, $r_s = 1/2$.\cite{Runeson2019} The Hamiltonian, Eqn.~\eqref{ODH}, and generally any Hermitian operator of electronic states, can be decomposed into the spin-operators and the identity\cite{Runeson2019, Bossion2021, Meyer1979} 
\begin{align}
    \hat{H} = H_0 \hat{\mathbb{I}} + H_x \hat{S}_x + H_y \hat{S}_y + H_z \hat{S}_z \textrm{,}
\end{align}
where the spin-operators $\hat{S}_i = \hat{\sigma}_i/2$ for $i \in \{x,y,z\}$ and the Pauli spin-matrices are
\begin{align} \label{Paulimatrices}
    \hat{\sigma}_x = \begin{bmatrix}
        0 & 1 \\ 1 & 0 
    \end{bmatrix} \ , \ \hat{\sigma}_y = \begin{bmatrix}
        0 & -i \\
        i & 0
    \end{bmatrix} \, \ \hat{\sigma}_z = \begin{bmatrix}
        1 & 0 \\ 0 & -1
    \end{bmatrix} \textrm{,}
\end{align}
such that 
\begin{align}
    H_0 = \frac{1}{2}\hat{\bo{P}}^\mathrm{T} \boldsymbol{\mu}^{-1}\hat{\bo{P}} + U(\hat{\bo{R}}) + \frac{1}{2}\mathrm{Tr}[\bo{V}(\hat{\bo{R}})] \textrm{,}
\end{align}
and 
\begin{align}
    \label{H}
    \bo{H} = \begin{bmatrix}
        H_x\\ H_y \\ H_z
    \end{bmatrix} = \begin{bmatrix}
        \Delta^* + \Delta\\
        i(\Delta^* - \Delta)\\ V_1 - V_2
    \end{bmatrix} \text{.}
\end{align}
Spin-mapping is therefore invariant to the initial choice of splitting of the potential matrix as $H_0$ will automatically include all state-independent terms.\cite{Runeson2019} It is easy to see that $H_x$ is twice the real part of the coupling ($\Delta$) and $H_y$ is twice the imaginary part. The classical phase-space is constructed by defining a spin-vector, ${\mathbf{\mathcal{S}}}$, which is the expectation value of the spin-operators in the spin-coherent states which are defined as\cite{radcliffePropertiesCoherentSpin1971}
\begin{align}
    \label{spin-coherent-states}
    \ket{\bo{u}} = \cos \frac{\theta}{2}e^{-i\phi/2} \ket{1} + \sin \frac{\theta}{2}e^{i\phi /2}\ket{2} \textrm{,}
\end{align}
where $\braket{\bo{u}}{\bo{u}}=1$ such that 
\begin{align}
    \mathcal{S}_i (\bo{u}) \equiv \bra{\bo{u}}\hat{S}_i\ket{\bo{u}} \textrm{,}
\end{align}
where
\begin{align} \label{S}
    {\mathbf{\mathcal{S}}}= \frac{1}{2}\begin{bmatrix}
        \sin\theta\cos\phi\\ \sin\theta\sin\phi \\ \cos\theta
    \end{bmatrix} \text{.}
\end{align}

For a Euclidean phase space, the quantum-mechanical trace in the correlation function, Eqn.~\eqref{General-CF}, is calculated using a Wigner transform\cite{wignerQuantumCorrectionThermodynamic1932, moyalQuantumMechanicsStatistical1949} 
\begin{align} \label{wigner}
    A_W(q, p) = \int_{-\infty}^{\infty} dy \, e^{-i p y} \left\langle q + \frac{y}{2} \middle| \hat{A} \middle| q - \frac{y}{2} \right\rangle \text{.}
\end{align}
which is calculated with classical-like integrals as the quantum operators are mapped onto a continuous phase-space. As the spin-vector is defined for two-states on the surface of the sphere, the Stratonovich-Weyl (SW) transform is used to map the operators onto a continuous function on the associated spin-mapping phase space.\cite{Runeson2019, Runeson2020, Bossion2021, bossionNonadiabaticMappingDynamics2022, brifPhasespaceFormulationQuantum1999, Stratonovich} First, the general operator, $\hat{A}$, is mapped to 
\begin{align}
    \hat{A} \to A_{Q}(\bo{u}) \equiv \bra{\bo{u}}\hat{A}\ket{\bo{u}} \textrm{,}
\end{align}
which is known as the Husimi or Q-function.\cite{Runeson2019, klimovGeneralizedSU2Covariant2017} This allows the definition of the quantum-mechanical trace as an integral over a classical phase-space 
\begin{align} \label{Q-function-trace}
    \mathrm{Tr} [\hat{A}] = \int \mathrm{d}\bo{u} \  A_{Q}(\bo{u}) = \frac{1}{2\pi}\int_0^{2\pi} \mathrm{d}\phi \int_0^\pi \sin \theta \ \mathrm{d} \theta \  A_{Q}(\bo{u}) \textrm{.}
\end{align}
Any operator can be decomposed into the spin-basis and the identity matrix via
\begin{align}
    \hat{A} = A_0 \mathbb{I} + \sum_i A_i \hat{\sigma}_i \textrm{,}
\end{align}
where
\begin{align}
    A_0 = \frac{1}{2}\mathrm{Tr}[\hat{A}\mathbb{I}] \quad , \quad A_i = \frac{1}{2}\mathrm{Tr}[\hat{A}\hat{\sigma}_i] \textrm{,}
\end{align}
and for simplicity, this is presented in terms of the Pauli spin-matrices. Hence
\begin{align} \label{A_q_sum}
    A_Q(\bo{u}) = A_0\braket{\bo{u}}{\bo{u}} + \sum_i A_i \braketthree{\bo{u}}{\hat{\sigma}_i}{\bo{u}} = A_0 + 2 \sum_i A_i \mathcal{S}_i \textrm{,}
\end{align}
such that evaluating the integrals in Eqn.~\eqref{Q-function-trace} results in the trace. However, this cannot be utilised to define the trace of two operators as 
\begin{align}
    \mathrm{Tr}[\hat{A}\hat{B}] \neq \int \mathrm{d}\bo{u} A_Q(\bo{u}) B_Q(\bo{u}) \textrm{,}
\end{align}
where $\bra{\bo{u}}\hat{A}\ket{\bo{u}}\bra{\bo{u}}\hat{B}\ket{\bo{u}} \neq \bra{\bo{u}}\hat{A}\hat{B}\ket{\bo{u}}$ unless $\hat{A}$ or $\hat{B}$ is the identity.\cite{Runeson2019} The SW kernel is defined as 
\begin{align}
    \hat{\omega}_s = \frac{1}{2}\mathbb{I} + 2 r_s \mathbf{\mathcal{S}} \cdot \hat{\boldsymbol{\sigma}} \textrm{,}
\end{align}
where it is easy to see that taking $r_s= 1/2$ allows Eqn.~\eqref{A_q_sum} to be written as
\begin{align} \label{alt-Q}
    A_s(\bo{u}) =  \mathrm{Tr}[\hat{A}\hat{\omega}_s] \textrm{.}
\end{align}
Hence, we can define three commonly used kernels where $s \in \{Q,P,W\}$ to have spin-radii of $1/2$, $3/2$ and $\sqrt{3}/2$ respectively for a two-level system.\cite{Runeson2019, Bossion2021, klimovGeneralizedSU2Covariant2017} The trace of a product of operators is 
\begin{align}
    \mathrm{Tr}[\hat{A}\hat{B}] = \int \mathrm{d}\bo{s} \  A_{s}(\bo{u}) B_{\bar{s}}(\bo{u}) \textrm{,}
\end{align}
where $s$ and $\bar{s}$ are referred to as \textit{dual} where $r_s \cdot r_{\bar{s}} = 3/4$.\cite{Runeson2019} This is required such that the prefactor arising from the integrals over the angles cancels with $r_s \cdot r_{\bar{s}}$.\cite{Runeson2019, Bossion2021} The P-function (Glauber–Sudarshan) and Q-function are dual and the W-function (Wigner) is self-dual.\cite{klimovGeneralizedSU2Covariant2017} By considering the case where $\hat{B}$ is the identity, Eqn.~\eqref{Q-function-trace} can also be shown for the P- and W-functions. The dual relations have been tabulated in Table~\ref{SW TABLE} alongside the corresponding ZPE parameter discussed later. The SW transform also preserves the identity operator in the electronic Hilbert space,\cite{bossionNonadiabaticMappingDynamics2022, Runeson2020} in contrast to the Wigner transform of the identity operator in the MMST mapping space, such that the total populations are always unity avoiding the ZPE-leakage in the mapping space that the MMST Hamiltonian suffers from.\cite{Runeson2020, sallerImprovedPopulationOperators2020, sallerIdentityIdentityOperator2019} 

\begin{table}[H]
\caption[Spin-radii, $r_s$, dual symbols, $\bar{s}$ and ZPE parameter, $\gamma$, for the three Q-, P- and W-functions.]{Spin-radii, $r_s$, dual symbols, $\bar{s}$ and ZPE parameter, $\gamma$, for the three Q-, P- and W-functions.  }
 \begin{center} 
    \begin{tabular}{p{0.15\linewidth}|p{0.15\linewidth}|p{0.15\linewidth}|p{0.15\linewidth}|p{0.15\linewidth}}
      \toprule 
      \textbf{$s$} & \textbf{ $\bar{s}$} & \textbf{ $r_s$ } & \textbf{ $r_{\bar{s}}$ } & \textbf{ $\gamma$ } \\ \hline
      \midrule 
      \rowcolor[HTML]{C0C0C0}
     Q & P & $1/2$ & $3/2$ & 0 \\
      P & Q & $3/2$&  $1/2$ & 2 \\
      \rowcolor[HTML]{C0C0C0} W & W & $\sqrt{3}/2$ & $\sqrt{3}/2$ & $\sqrt{3}-1$ \\
      \bottomrule 
    \end{tabular}
    \label{SW TABLE}
\end{center}
\end{table}

Considering the Q-relation on the spin-operator\cite{Bossion2021} 
\begin{align}
    [\hat{S}_i]_{Q} (\bo{u}) = \bra{\bo{u}}\hat{S}_i\ket{\bo{u}} = \mathcal{S}_i \textrm{,}
\end{align}
such that the spin-operator is mapped onto the spin-vector. This is equivalent to
\begin{align}
    [\hat{S}_i]_{s} (\bo{u})  = \mathrm{Tr}\left[ \hat{S}_i \hat{\omega}_s \right] = 2r_s\mathcal{S}_i \textrm{,}
\end{align}
for the general SW kernel.
The spin-mapping Hamiltonian is therefore\cite{Runeson2019} 
\begin{align}
    \label{spin-hamiltonian}
    H_{\textrm{SM}} (\bo{u}) = \frac{1}{2} \bo{P}^\textrm{T}\boldsymbol{\mu}^{-1}\bo{P} + U (\bo{R}) + \frac{1}{2} \mathrm{Tr}[\bo{V}(\bo{R})] + \frac{1}{2}\bo{H}(\bo{R})\cdot \bo{u} \text{,}
\end{align}
where the operators have been replaced with the classical variables and for notational simplicity, the dependence on $\bo{R}$ is dropped from here. We define our non-canonical scaled spin-vector as 
\begin{align}
    \label{u}
    \bo{u} =  \begin{bmatrix}
        u_x\\ u_y \\ u_z
    \end{bmatrix} = 4r_s \begin{bmatrix}
        S_x\\ S_y \\ S_z
    \end{bmatrix} = 2r_s\begin{bmatrix}
        \sin\theta\cos\phi\\ \sin\theta\sin\phi \\ \cos\theta
    \end{bmatrix} \text{,}
\end{align}
which encodes the $r_s$ dependence of the Hamiltonian and the elements of $\mathbf{\mathcal{S}}$ are the spin-vector in Eqn.~\eqref{S}.\cite{Runeson2019} The scaled spin-vector is non-canonical and is propagated using Heisenberg's EOM\cite{Runeson2019, Bossion2021}      
\begin{align}
    \label{Heisenberg}
    \dot{\bo{u}} = \bo{H} \times \bo{u}\text{,}
\end{align}
where it is also interesting to note that the elements of $\bo{H}$ can be calculated by 
\begin{align}
    H_i = \mathrm{Tr}[\bo{V}\sigma_i] \text{.}
\end{align}
The spin-mapping Hamiltonian is then,
\begin{align}
    \label{spin-ham-full}
    H_{\textrm{SM}} (\bo{u}) = \frac{P^2}{2m} + U + \frac{1}{2} \mathrm{Tr}[\bo{V}] + \frac{1}{2}\left[ 2 \Re(\Delta) u_x + 2 \Im(\Delta) u_y+ (V_1 - V_2)u_z \right] \text{.}
\end{align}
Spin-mapping has been successfully extended to more than two electronic states by extending to the $SU(F)$ Lie group using the Generalised Gell Mann (GGM) matrices.\cite{Runeson2020, bossionNonadiabaticMappingDynamics2022, bossionNonadiabaticRingPolymer2023} Whilst the spin-vector is $F^2-1$ dimensional, the spin-mapping system is defined by $2F-2$ angles as in Eqn.~\eqref{S}, thus removing two redundant dof compared to the MMST mapping.

\subsection{Connection to the MMST Hamiltonian}
For the MMST mapping, the electronic states are mapped onto singly-excited oscillators (SEO) using the creation and annihilation operators as\cite{Stock1997, Hele2016}
\begin{align}\label{creation_annihaliation}
    \ket{n}\bra{m} = \hat{a}_n^\dagger \hat{a}_m  \textrm{,}  
\end{align}
where $\hat{a}_m = (\hat{q}_m + i \hat{p}_m)/\sqrt{2}$ and $\dagger$ is the conjugate transpose. Several approaches can be used to obtain the classical limit,\cite{Stock1997, Hele2016} such that the classical MMST mapping Hamiltonian in the diabatic representation is\cite{Meyer1979, Stock1997}
\begin{align} \label{MMST-imag-notsplit}
    H_{\textrm{MMST}} = \frac{1}{2} \bo{P}^\textrm{T}\boldsymbol{\mu}^{-1}\bo{P} + U(\bo{R}) + \frac{1}{2}\left\{ \bo{p}^\textrm{T}\bo{V}(\bo{R})\bo{p} + \bo{q}^\textrm{T}\bo{V}(\bo{R})\bo{q} - \gamma \textrm{Tr}[\bo{V}(\bo{R})]\right\} \text{,}
\end{align}
where $\bo{R}$ and $\bo{P}$ are $K$ dimensional vectors of nuclear position and momenta, and the mapping variables, $\bo{q}$ and $\bo{p}$, are $F$ dimensional vectors of electronic position and momenta. The ZPE parameter, $\gamma$, is unity in the standard MMST mapping.\cite{Muller1999}

The spin-mapping Hamiltonian, Eqn.~\eqref{spin-ham-full}, is equivalent to the ZPE-Reduced MMST Hamiltonian, where $\gamma = 2r_s -1$, for the following coordinate transform for a two-state system\cite{Runeson2019} 
\begin{subequations} \label{mmst-spin}
    \begin{align}
        u_x &= q_1q_2 + p_1p_2 \text{,} \\
        u_y &= q_1p_2 - q_2p_1 \text{,}\\
        u_z &= \tfrac{1}{2}(q_1^2 + p_1^2 - q_2^2 -p_2^2) \text{,} 
    \end{align}
\end{subequations}
where $\bo{q}$ and $\bo{p}$ are canonical variables and the spin-radius is, 
\begin{align} \label{radius-mmst}
    4r_s = q_1^2 + p_1^2 + q_2^2 +p_2^2 \text{.}
\end{align}

\subsection{Matsubara Dynamics}
Here, we discuss how the Matsubara dynamics conserves the QBD for a single surface.\cite{Hele2015} We note that this approach has been utilised for the nuclear variables in the NA-Mats and SM-NA-Mats methods.\cite{Chowdhury2021, bossionNonadiabaticRingPolymer2023}

The Kubo-transformed (KT)-CF 
\begin{align} \label{kubotransformed}
    C_{AB}(t) = \frac{1}{Z \beta} \int^{\beta}_{0} d\lambda \text{Tr} \left[ e^{-(\beta-\lambda)\hat{H}} \hat{A} e^{-\lambda\hat{H}} e^{i\hat{H}t} \hat{B} e^{-i\hat{H}t} \right]
\end{align}
has a higher degree of symmetry compared to the standard, unsymmetrized CF in Eqn.~\eqref{General-CF} and the two are related through a series of Fourier transforms.\cite{kuboFluctuationdissipationTheorem1966, Craig2004, Willatt2017} 

Matsubara dynamics approximates a discretised imaginary-time path-integral version of the KT-CF through truncation of the Liouvillian, the generalised KT-CF,\cite{Hele2015, Hele2013, Hele2013b, Althorpe2013, cookElectronicPathIntegral2025} 
\begin{align}
    \label{generalised-KT-cf}
    C^{[N]}_{AB}(t) 
    &=  \frac{1}{Z} \prod^{N}_{j} \iint d{\bf{R}}_{j} d{\boldsymbol{\mu}}_{j} \nonumber \\ & \quad \times \biggl\langle {R_{j-1} - \mu_{j-1}/2} \bigg\vert {\frac{1}{2}(\hat{A}e^{-\beta_N\hat{H}} + e^{-\beta_N\hat{H}}\hat{A})} \bigg\vert  R_j + {\mu}_j/2 \biggr\rangle \nonumber \\ & \quad \times \braketthree{{R}_j + {\mu}_j/2} {e^{i\hat{H}t} \hat{B} e^{-i\hat{H}t}}{R_j - \mu_j/2} \text{,}
\end{align}
where the operators act once in the path-integral such that $\hat{A} = (\sum_j^N \hat{A}_j)/N$ and likewise for $\hat{B}$, the indices are cyclic ($R_N =R_0$) and $\beta_N= \beta/N$.
If the operators can be written as a linear sum of operators that act once in the path-integral, as is defined above, the generalised KT-CF becomes equivalent to the KT-CF in the limit of $N \to \infty$.\cite{Hele2013, Willatt2017, Hele2015} For clarity, non-linear operators under this definition would include the flux and side operators.\cite{Althorpe2013, Hele2013, Hele2013b, Hele2014} 


In practice, one computes the Wigner transformed generalised KT-CF\cite{Hele2016, wignerQuantumCorrectionThermodynamic1932,hilleryDistributionFunctionsPhysics1984} 
\begin{align} \label{GKT_wigner-nuc}
    C^{[N]}_{AB}(t) &= \frac{1}{Z(2\pi)^{N} }\iint d{\bf{R}} d{\bf{P}} [e^{-\beta_N \hat{H}}\hat{A}]_{\overline{W}} \times [\hat{B}(t)]_W \text{,}
\end{align}
where the real-time and imaginary-time terms are defined as 
\begin{subequations} \label{wigner-rt-it-nuc}
\begin{align}
    &[\hat{B}(t)]_W = \prod^{N}_{j} \int d{\mu}'_{j} e^{i{\boldsymbol{\mu}}'_j {P}_j}  \braketthree{{R}_j + \mu'_j/2} { \hat{B}(t) }{R_j - \mu'_j/2} \text{,}\\ 
    &[e^{-\beta_N \hat{H}}\hat{A}]_{\overline{W}} = \prod^{N}_{j} \int d{\mu}_{j}  e^{i{\mu}_j {P_j}} \left. \biggl\langle {R_{j-1} - \mu_{j-1}/2} \bigg\vert \mathcal{A} \middle\vert  R_j + {\mu}_j/2 \right\rangle \text{,}
\end{align}
\end{subequations}
where $\mathcal{A} = (\hat{A}e^{-\beta_N\hat{H}} + e^{-\beta_N\hat{H}}\hat{A})/2$ is the symmetric operator, $\hat{B}(t) = e^{i\hat{H}t} \hat{B} e^{-i\hat{H}t}$ is the time-evolved operator and, $\int d\bo{R} = \prod_j^N \int d \bo{R}_j$ and likewise for $\bo{P}$ and $\mu$.\cite{Hele2015, Hele2016} The form of $[\cdot]_{\overline{W}}$ and $[\cdot]_W$ are different, where the latter is a simple bead-average of the Wigner transform and the former is more complex coupling variables in $j$ and $j+1$.\cite{Hele2015, cookElectronicPathIntegral2025} 

Matsubara dynamics transforms into the free ring-polymer normal modes  
\begin{align} \label{normalmodetransformation}
    \check{R}_{k} = \sum_{j=1}^N T_{jk}R_{j} \text{,}
\end{align}
where $j$ and $k$ are the bead and mode indices respectively. The transformation matrix, $\bf{T}$, for an even number of beads is 
\begin{align}
    \label{transformation}
    T_{jk} = \begin{cases}
        \sqrt{\frac{1}{N}} \quad & k=0 \\
        \sqrt{\frac{2}{N}} \cos(2\pi jk/N) \quad &1 \leq k \leq N/2-1 \\
        \sqrt{\frac{1}{N}}(-1)^j \quad & k=N/2 \\
        \sqrt{\frac{2}{N}} \sin(2\pi jk/N) \quad & N/2-1 \leq k \leq N-1 \text{,} 
    \end{cases} 
\end{align}
which is equivalent to the half-complex Fourier transform. For odd numbers of beads, the transformation matrix is the same as above but the $k=N/2$ term is removed and can be seen in Ref.~\hbox{[\!\citenum{Hele2015}]}. The ring-polymer has springs connecting adjacent beads, which are not explicitly present in the Matsubara Hamiltonian, for which the transformation matrix was chosen to diagonalise the spring constant matrix.\cite{ceriottiEfficientStochasticThermostatting2010, Hele2015, cookElectronicPathIntegral2025} 
The frequencies of the ring-polymer normal modes are \cite{Hele2015}
\begin{align}
\label{normalmodefre}
\omega_k = \frac{2}{\beta_N}\sin \left(\frac{k \pi}{N}\right) \text{,}
\end{align}
and in the limit $N \to \infty$ and $M\ll N$, the frequencies of the $M$ lowest normal modes tend to the `Matsubara' frequencies 
\begin{align}
    \label{mats-freq}
    \bar{\omega}_k = \lim_{N \to \infty} \omega_k = \frac{2k \pi}{\beta} \text{.}
\end{align}
To ensure convergence with respect to $N$, the Matsubara modes are defined as
\begin{align}
    \label{matsubara-modes}
    \bar{R}_k = \lim_{N \to \infty} \frac{\check{R}_k}{\sqrt{N}} \text{,}
\end{align}
for the lowest $M$ modes and $\bar{R}_0$ is the centroid.\cite{Hele2015, heleCommunicationRelationCentroid2015} Matsubara dynamics is not the first approach based on Matsubara modes as, for many years, they have been utilised for computing equilibrium properties with path-integrals.\cite{Ceperley1995, chakravartyPathIntegralSimulations1997, chakravartyComparisonEfficiencyFourier1998,freemanMonteCarloMethod1984}  


Any superposition of the Matsubara modes results in a smooth and differentiable distribution in imaginary-time.\cite{Hele2015} Matsubara dynamics approximates the Liouvillian to be a function of only the lowest normal modes through truncation in the higher normal modes, which assumes that the propagation of the higher and lower normal modes can be decoupled, and surprisingly results in classical dynamics (without the need to explicitly truncate the Liouvillian in powers of $\hbar$).\cite{Hele2015} The higher, unpropagated normal modes are constrained by the QBD such that  they can be integrated out which results in QBD conserving dynamics (as the remaining lowest path-integral modes are a smooth function of imaginary-time).\cite{Hele2015, goldsteinClassicalMechanics1980} Additionally, many observables can be written as a function of only a finite number of the lowest normal modes of the nuclear position and/or momenta.\cite{Hele2015, heleAlternativeDerivationRingpolymer2016} However, Matsubara dynamics is challenging to numerically evaluate as it possesses a highly oscillatory phase factor which introduces a `sign' problem such that a large number of trajectories is required to converge the initial quantum statistics.\cite{Hele2015,Ceperley1995}

There are two previously proposed nonadiabatic Matsubara methods which use the Matsubara approach for the nuclear dof, but leave the electronic dof unchanged either in the MMST mapping representation (NA-Mats),\cite{Chowdhury2021} or the spin-mapping representation (SM-NA-Mats).\cite{bossionNonadiabaticRingPolymer2023} Both methods generally do not conserve the QBD.\cite{Chowdhury2021, bossionNonadiabaticRingPolymer2023} In addition to the Matsubara dynamics approximations, the back action from the electronic dof onto the nuclear subsystem has been ignored and the full effects of this approximation has yet to be explored.\cite{Chowdhury2021, bossionNonadiabaticRingPolymer2023} In order to calculate a correlation function with $N$ electronic beads and $M$ nuclear Matsubara modes, a back-transform to obtain $N$ nuclear beads from only the Matsubara modes is introduced and the effects of this have not been determined.\cite{Chowdhury2021, bossionNonadiabaticRingPolymer2023} 
There are two known cases where the proposed NA-Mats method conserves the QBD where the electronic and nuclear dof are decoupled, including two limits of an electronically adiabatic system and an electronic-only system, and when the electronic dof are linearly coupled to a harmonic bath.\cite{Chowdhury2021, cookElectronicPathIntegral2025} A discussion of such cases has not, to our knowledge, been published for SM-NA-Mats although we expect it to be similar.\cite{bossionNonadiabaticRingPolymer2023} In addition to the `sign' problem arising from the Matsubara phase, both methods also introduce a second `sign' problem for a large number of electronic states or beads due to a complex factor in the electronic part of the CF.\cite{Chowdhury2021, bossionNonadiabaticRingPolymer2023}

This does not mean that there does not exist a nonadiabatic Matsubara method which conserves the QBD, and we suspect that to achieve conservation of the QBD, a similar approach to Matsubara dynamics for the normal modes of the electronic dof may be required. When investigating one choice of electronic metric, the MMST variables, the following desirable criteria were proposed,\cite{cookElectronicPathIntegral2025}
\begin{enumerate} [itemsep=0pt]
    \item Observables are a function of a finite number of the lowest normal modes,
    \item The distribution of the normal modes narrows as they increase in frequency (such that they can be integrated out when dynamics is truncated),
    \item Truncating the normal modes results in conservation of the QBD,
    \item Truncating the normal modes provides a good approximation to exact (untruncated) dynamics. 
\end{enumerate} 
The MMST path-integral normal modes were not found to fulfil these metric criteria, probably due to the lack of a constraint on the MMST electronic normal modes in contrast to the springs between beads for nuclear ring-polymer normal modes, and that the observables are in general a function of all normal modes such that truncating in the normal modes led to inaccurate dynamics.\cite{cookElectronicPathIntegral2025} This may explain why, to the best of our knowledge and despite decades of research, there does not yet exist a MMST mapping method which conserves the distribution and reproduces Rabi oscillations simultaneously for a general system.
However, this does not mean an alternative metric will not satisfy these desirable properties. Due to the successes with spin-mapping approaches,\cite{Runeson2019, Bossion2021, MASH, Amati2023} we investigate the electronic path-integral normal modes of the spin-mapping representation in this work.

\section{Methodology} \label{methodology}

We utilise Fermi's golden rule to determine if a spin-mapping electronic metric will be constrained by the QBD. After defining the spin-mapping normal modes, we then simplify the problem by considering an electronic-only system to investigate if observables are a function of a finite number of the lowest spin-mapping normal modes and if truncating in these normal modes results in QBD conservation and good quality dynamics. We believe it is highly likely that if these properties are not present for an electronic-only system, they will also not be present for a coupled nuclear-electronic system.\cite{cookElectronicPathIntegral2025} 

\subsection{Boltzmann Penalty} \label{scaling-of-the-boltzmann-factor-with-N}

We turn our attention to considering what will act as a `penalty' or constraint on the electronic normal modes. For nuclear dof in conventional RPMD,\cite{Craig2004} the Boltzmann penalty applies to constrain high-frequency motion of the imaginary-time path-integral, as seen from the ring-polymer frequencies in Eqn.~\eqref{normalmodefre} and in the distribution of a free ring-polymer in a harmonic potential. Here, we explore how the Boltzmann distribution may penalise various configurations of \emph{electronic} dof. We acknowledge that the arguments in this section are presented somewhat heuristically, though we later provide strong numerical evidence for them, and we leave a full analysis of the electronic constraint arising from the imaginary-time electronic Boltzmann distribution as future research. 

When the coupling between states is small, i.e.\ $\beta \Delta\ll 1$, as is usually the case for nonadiabatic dynamics, Fermi's golden rule applies. This states that the rate of the transition between two states is proportional to the coupling squared, $k \propto \Delta^2$.\cite{lawrenceImprovedPathintegralMethod2020, nitzanNitzanChemicalDynamics2024} Hence, the `penalty' (or amplitude) of this transition is proportional to the coupling. For two-states, this is the off-diagonal elements of the potential matrix. As we are considering a path-integral formalism to obtain normal modes, the penalty incurred `hopping' between states can be determined by investigating the scaling of $e^{-\beta_{N}{\bf{V}}}$ with \textit{N} beads where $\beta_{N} = \beta/N$.

If we consider a symmetric potential matrix as in Eqn.~\eqref{diabatic pot mat}, where $\Delta$ is real, which can be decomposed into traceless and identity-like components
\begin{align} \label{V_traceless}
\mathbf{V} = \frac{V_1+V_2}{2}\mathbb{I}  + \begin{bmatrix} \frac{V_1-V_2}{2} &\Delta \\ \Delta & \frac{V_2-V_1}{2}
\end{bmatrix}  \text{,}
\end{align}
the exponent can then be represented through Taylor expansion as 
\begin{align}
    e^{-\beta_N\bo{V}} = e^{-\beta_N V_0} \left[ \cosh(\beta_N \delta ) \mathbb{I} -  \frac{\sinh (\beta_N \delta )}{\delta} \bar{\bo{V}}\right] \text{,}
\end{align}
where $\delta = \sqrt{(V_1 - V_2)^2/4 + \Delta^2}$ and $V_0$ and $\bar{\bo{V}}$ are the identity and traceless component of $\bo{V}$ respectively, such that
\begin{align}
    e^{-\beta_N\bo{V}} = e^{-{\beta_N V_0}} \begin{bmatrix}
        \cosh(\beta_N \delta )  -  \frac{(V_1-V_2)\sinh (\beta_N \delta )}{2\delta} & -  \frac{\Delta\sinh (\beta_N \delta )}{\delta} \\
         -  \frac{\Delta\sinh (\beta_N \delta )}{\delta} & \cosh(\beta_N \delta )  +  \frac{(V_1-V_2)\sinh (\beta_N \delta )}{2\delta} 
    \end{bmatrix} \text{,}
\end{align}
which is exact for any $N$. Taking the limit as $N\to \infty$ uses the following
\begin{subequations}
    \begin{align}
        \lim_{N\to \infty} e^{-\beta_N V_0} &\to 1 {-\beta_N V_0} \text{,}\\
        \lim_{N\to \infty}  \sinh (\beta_N \delta ) & \to \beta_N \delta \text{,}\\
        \lim_{N\to \infty}  \cosh (\beta_N \delta ) & \to  1 +(\beta_N \delta)^2/2 \text{,}
    \end{align}
\end{subequations}
where we have truncated higher orders of $\beta_N$ as this becomes very small as $N \to \infty$. Hence 
\begin{subequations}
    \begin{align}
        \lim_{N\to \infty}   e^{-\beta_N\bo{V}} &= \left\{1 -\beta_N V_0 \right\} \begin{bmatrix}
         1 - \frac{(V_1-V_2)\beta_N }{2}  & -  \beta_N \Delta \\
         -  \beta_N \Delta &1 + \frac{(V_1-V_2)\beta_N }{2}
    \end{bmatrix} + \mathcal{O}(\beta_N^2)\\
    &= \begin{bmatrix}
         1 - \beta_N V_1 & -  \beta_N \Delta \\
         -  \beta_N \Delta & 1 - \beta_N V_2
    \end{bmatrix} + \mathcal{O}(\beta_N^2) \text{.}
    \end{align}
\end{subequations}

\begin{figure}
    \centering
    \includegraphics[width=0.8\linewidth]{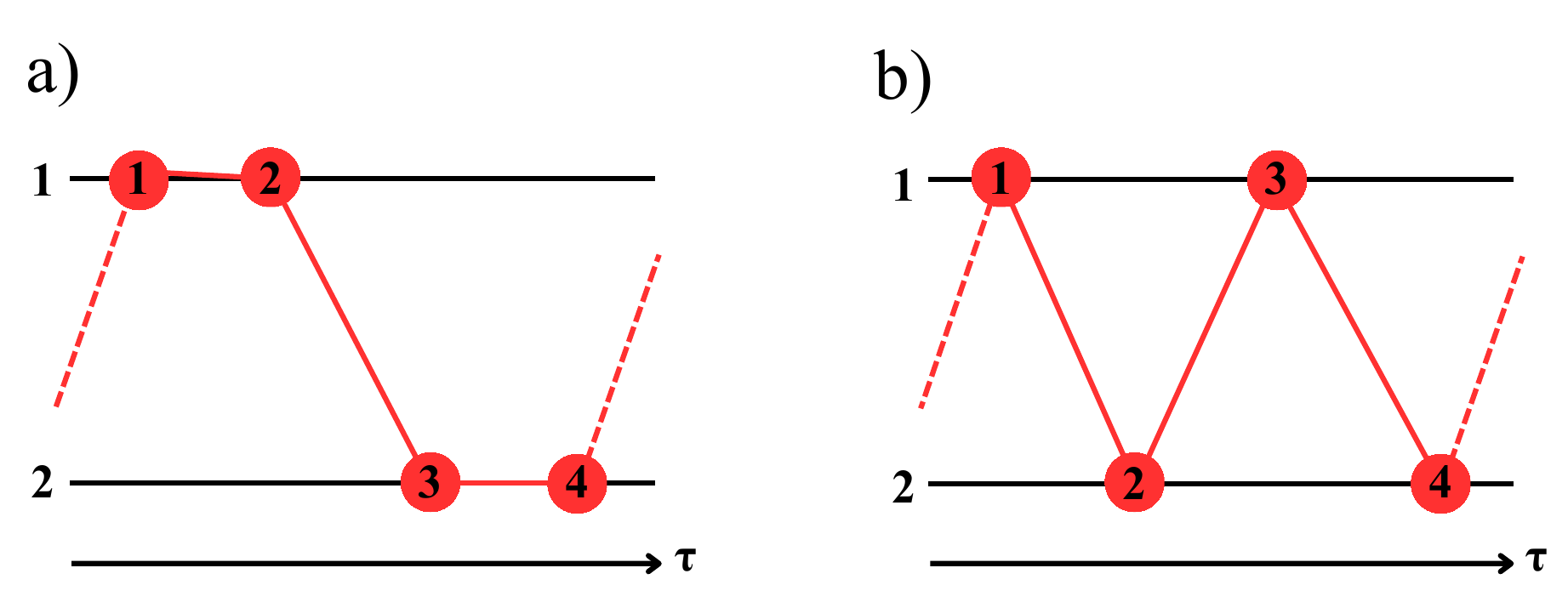}
    \caption{\label{fig:weight}Schematic for a) two crossing and b) four crossings between the two electronic states using four beads in imaginary-time ($\tau$). Due to the Boltzmann penalty, one would expect that b) is less favourable than a).}
\end{figure}

Therefore, a `hop' between states carries a penalty proportional to the off-diagonal elements. In the limit of the number of beads tending to infinity, where the dynamics approaches exact imaginary-time dynamics, this is proportional to $\beta \Delta /N $. We will refer to this as the Boltzmann penalty. For a path-integral where there is a `hop' between every bead, shown in Figure~\ref{fig:weight}b for four beads, the penalty would be proportional to $(\beta \Delta /N)^N$. This is less favourable than Figure~\ref{fig:weight}a where there are only two `hops' between states. 


This implies that an electronic metric that is constrained by the QBD is the \emph{difference} in electronic population between the states, which is twice the $z$-component of the spin-mapping vector for two states. This is easily seen through the transformation to the equivalent MMST Cartesian variables in Eqn.~\eqref{mmst-spin}, where the $z$-component becomes $q_{1j}^2 + p_{1j}^2 - q_{2j}^2 -p_{2j}^2$. Hence, it may be possible to derive a nonadiabatic dynamics method that satisfies all the criteria using the spin-mapping representation of nonadiabatic dynamics. 

\subsection{Spin-Mapping Normal Modes} \label{spin-mapping-normal-modes}

In the two-level case, the scaled spin-vector can be defined as Eqn.~\eqref{u} which transforms into MMST variables as Eqn.~\eqref{mmst-spin} with a radius of Eqn.~\eqref{radius-mmst}. As in the previous electronic normal modes study,\cite{cookElectronicPathIntegral2025} an electronic-only system is considered such that the Hamiltonian in the spin-mapping representation is 
\begin{align}
    H &= \frac{1}{2}\mathrm{Tr}\left[\bo{V}\right] + \frac{1}{2} \bo{H}\cdot{ \bo{u}} \textrm{,}
\end{align}
and the potential matrix, $\bo{V}$, is of the form seen in Eqn.~\eqref{diabatic pot mat} but is independent of nuclear position. The Hamiltonian vector is defined in Eqn.~\eqref{H}.  
The electronic EOM then follow Heisenberg's EOM as in Eqn.~\eqref{Heisenberg} 
\begin{subequations} \label{u-heisenberg}
    \begin{align} 
        \dot{u}_x &= u_z H_y -u_y H_z \text{,} \\
        \dot{u}_y &= -u_z H_x + u_xH_z \text{,}\\
        \dot{u}_z &= u_y H_x - u_x H_y \text{.}
    \end{align}
\end{subequations} 
We utilise the recently developed Spin-MInt algorithm's electronic evolution where the cross product in Eqn.~\eqref{Heisenberg} is recast as a matrix multiplication\cite{cookSpinMIntAlgorithmAccurate2026}
\begin{subequations} \label{W-evolution}
    \begin{align}
        \dot{\bo{u}} &= \begin{bmatrix}
            0 & -H_z & H_y \\
            H_z & 0 & -H_x \\
            -H_y & H_x & 0 
        \end{bmatrix} \begin{bmatrix}
            u_x \\
            u_y\\
            u_z
        \end{bmatrix} \\
        &= -i \begin{bmatrix}
            0 & -i(V_1 - V_2) & \Delta-\Delta^* \\
        i(V_1-V_2) & 0 & -i(\Delta^* + \Delta) \\
            -(\Delta-\Delta^*) & i(\Delta^* + \Delta) & 0 
        \end{bmatrix} \begin{bmatrix}
            u_x \\
            u_y\\
            u_z
        \end{bmatrix} \\
        &= -i\bo{W}\bo{u} \text{,}
    \end{align}
\end{subequations}
where $\bo{W}$ is Hermitian, skew-symmetric, and all elements are imaginary or zero. 
Integration results in 
\begin{align}
    \label{u-prop}
    \bo{u}(t+\Delta t) = e^{-i\bo{W}\Delta t}\bo{u}(t) \text{,}
\end{align} 
which can be evaluated computationally using the decomposition of $\bo{W}$ into eigenvalues and eigenvectors.\cite{cookSpinMIntAlgorithmAccurate2026}

For $N$ beads, one can define a spin-matrix where the columns are the $N$ spin-vectors as
\begin{align}
{\bf{U}} = \begin{bmatrix}
     {\bf{u}}_{x} \\
     {\bf{u}}_{y} \\
     {\bf{u}}_{z}
    \end{bmatrix} = \begin{bmatrix} u_{x,1} &
     u_{x,2} &
     \hdots &
     u_{x,N} \\
     u_{y,1} &
     u_{y,2} &
     \hdots &
     u_{y,N}\\
     u_{z,1} &
     u_{z,2} &
     \hdots &
     u_{z,N}
    \end{bmatrix} \textrm{.}
\end{align}




As we consider an electronic-only system, the bead propagation is separable as ${\bf{W}}$ is the same for all beads such that\cite{cookElectronicPathIntegral2025}
\begin{align} \label{u-beads-prop}
    \bo{U}_j(t) = e^{-i {\bf{W}} t} \bo{U}_j(0) \textrm{,}
 \end{align}
where $\bo{U}_j$ is the $j$-th column of $\bo{U}$ which is the spin-vector for the $j$-th bead. The normal modes follow Eqn~\eqref{normalmodetransformation} such that for ${\bf{u}}_z$\cite{cookElectronicPathIntegral2025, Hele2015} 
\begin{align}
    \check{u}_{z,k} = \sum_j^N T_{jk}u_{z,j} \textrm{,}
\end{align}
and likewise for ${\bf
u}_x$ and ${\bf{u}}_y$ such that
\begin{align}
\check{{\bf{U}}} = \begin{bmatrix}
     \check{{\bf{u}}}_{x} \\
     \check{{\bf{u}}}_{y} \\
     {\check{\bf{u}}}_{z}
    \end{bmatrix} \textrm{,}
\end{align}
where we use the same transformation matrix as for a free ring-polymer which is an orthonormal matrix defined as Eqn.~\eqref{transformation}.\cite{ceriottiEfficientStochasticThermostatting2010}
The lowest (zeroth) normal mode of ${\bf{u}}_z$ is therefore defined as\cite{Hele2015} 
\begin{align} \label{u-z-lowest}
    \check{u}_{z,0} =  \frac{1}{\sqrt{N}} \sum_j^N{u_{z,j}} \textrm{,}
\end{align}
and likewise for ${\bf{u}}_x$  and  ${\bf{u}}_y$, which is related to the centroid defined as
\begin{align} \label{u-z-centroid}
    \bar{u}_{z} =  \frac{1}{N} \sum_j^N{u_{z,j}} = \frac{1}{\sqrt{N}}\check{u}_{z,0} \textrm{,}
\end{align}
and likewise for ${\bf{u}}_x$  and  ${\bf{u}}_y$ where the additional factor of $1/\sqrt{N}$ ensures convergence.\cite{Hele2015} 
The normal mode propagation is then 
\begin{subequations} \label{eq:ddtunk}
\begin{align}
    \frac{\mathrm{d}}{\mathrm{d} t}\check{u}_{n, k} &= \sum_j ^N T_{jk}\dot{u}_{n, j}(t) \\ 
    &= \sum_j^N \sum_{m}^3 -i {\bf{W}}_{nm}T_{jk}u_{m, j}(t) \\
    &= \sum_{j,r}^N \sum_{m}^3 -i {\bf{W}}_{nm}T_{jk}T_{jr}\check{u}_{m,r}(t) \textrm{,}
    \end{align}
\end{subequations}
where $x, y, z $ have been generalised to $n$ and $m$ and the following back-transform is utilised 
\begin{align}
    u_{m,j} = \sum_{r}^N T_{jr}\check{u}_{m,r} \textrm{.}
\end{align}
Due to no nuclear dependence of the potential matrix (and therefore of $\bo{W}$) and as the transformation matrix is orthonormal we can recast Eqn.~\eqref{eq:ddtunk} as
\begin{align}
    \frac{\mathrm{d}}{\mathrm{d} t} \check{\bo{U}}_{k} 
    &= -i {\bf{W}}\check{\bo{U}}_{k}(t) \textrm{,}
\end{align}
which integrates as in Eqn.~\eqref{u-beads-prop}.
Hence, the normal mode propagation of $\check{\bo{U}}$ is also separable in the case where there is no nuclear dependence of $\bo{W}$. 

\subsection{Spin-Mapping Correlation Function} \label{SM-CF}

We will now derive the generalised KT-CF entirely in the spin-mapping representation. Using the SW transform as in Eqn.~\eqref{alt-Q} for a general kernel, $s$, and following Ref.~[\!\citenum{bossionNonadiabaticRingPolymer2023}], the generalised KT-CF becomes
\begin{subequations}
    \begin{align}
        C_{AB}^{[N]}(t) = \frac{1}{Z} \int [e^{-\beta_N \bo{V}}\hat{A}]_{\bar{s}} \times [\hat{B}(t)]_{s} \ \mathrm{d}\{\bo{U}_j\} \textrm{,}
    \end{align}
\end{subequations}
where $\mathrm{d}\{\bo{U}_j\} = \prod_j^N \mathrm{d}\bo{U}_j$ and
\begin{align}
    \int \mathrm{d}\bo{U}_j = \frac{1}{2\pi}\int_0^\pi \sin \theta_j \ \mathrm{d} \theta_j \int_0^{2\pi} \mathrm{d}\phi_j  \textrm{.}
\end{align}
The imaginary-time term is
\begin{align} \label{imag-term}
    [e^{-\beta_N \bo{V}}\hat{A}]_{\bar{s}} = e^{-\beta V_0} \mathrm{Tr} \left[ \prod_j^N \mathcal{A}_j \hat{\omega}_{\bar{s}j}\right] \textrm{,}
\end{align}
where $\bo{V} = V_0 \mathbb{I} + \bo{H}\cdot\boldsymbol{\sigma}/2$ with $V_0 = \mathrm{Tr}[\bo{V}]/2$. The symmetrized operator is $\mathcal{A}_j = (e^{-\beta_N \bo{H} \cdot \boldsymbol{\hat{\sigma}}/2} \hat{A}_j + \hat{A}_j e^{-\beta_N \bo{H} \cdot \boldsymbol{\hat{\sigma}}/2})/2$ and as $\bo{V}$ is the same for each bead, $\left( e^{-\beta_N V_0} \right)^N = e^{-\beta V_0}$. The SW kernel is 
\begin{align}
    \hat{\omega}_{sj} = \frac{1}{2} \left( \mathbb{I} + \bo{U}_j \cdot \boldsymbol{\hat{\sigma}} \right) \textrm{,}
\end{align}
which is defined for both the dual-indices as in Section~\ref{spin-mapping-BT} where we let $\bo{U}_j= 4r_s \mathbf{\mathcal{S}}_j$ for the $j$-th bead. The real-time propagation is
\begin{subequations}
\begin{align}
    [\hat{B}(t)]_{s} &= \frac{1}{N}\sum_{j}^N  \mathrm{Tr}[\hat{B}_j \hat{\omega}_{sj}(t)] \\
    &= \frac{1}{N}\sum_{j}^N B_{j,0} + \bo{B}_{j}\cdot \bo{U}_j(t) \textrm{,}
\end{align}
\end{subequations}
where $B_{j,0} = \mathrm{Tr}[\hat{B}_j]/2$ and $B_{j,i} = \mathrm{Tr}[\hat{B}_j \hat{\sigma}_i]/2$ which are time-independent. Note that we have chosen to time-evolve the spin-vector but an equivalent approach would be instead time-evolving the operator. Hence, the difference in state populations is 
\begin{subequations}
    \begin{align}
        [\ketbra{1}{1}(t)-\ketbra{2}{2}(t)]_{s} &= \frac{1}{N}\sum_{j}^N \left[\frac{1}{2} + \frac{1}{2} u_{z,j}(t) - \frac{1}{2} + \frac{1}{2} u_{z,j}(t)\right] =  \bar{u}_{z}(t)
        \textrm{,}
    \end{align}
\end{subequations}
as the population of states is
\begin{subequations}
    \begin{align}
        [\ketbra{1}{1}(t)]_{s} 
        &= \frac{1}{2} \left[ 1 + \bar{u}_{z} (t)\right] \\
        [\ketbra{2}{2}(t)]_{s} 
        &= \frac{1}{2} \left[ 1 - \bar{u}_{z} (t)\right]
        \textrm{,}
    \end{align}
\end{subequations}
which are all a function of only the centroid of $\bf{U}$, Eqn.~\eqref{u-z-centroid} and it is quite easy to see that in general linear operators will result in a constant plus a linear combination of $\bar{u}_{x}$, $\bar{u}_{y}$ and $\bar{u}_{z}$. The sampling and partition function for this CF is outlined in Appendix~\ref{sampling-elec}.

We therefore see that spin-mapping fulfils the desirable metric criteria that the observable is a function of a finite number of the lowest normal modes (in this case only the centroid) and the Boltzmann penalty suggests that the higher normal modes will be constrained by the QBD. These two properties were not obtained when investigating MMST normal modes.\cite{cookElectronicPathIntegral2025}

\subsection{Conservation of the Quantum Boltzmann Distribution} \label{cons-sm}
To prove that the distribution is conserved in this representation, the electronic-only Liouvillian needs to be redefined. As the spin-vector is non-canonical, a transformation into conjugate canonical variables is required. For two states, these are the angles that are linked to the spin-vector through Eqn.~\eqref{u}\cite{bossionNonadiabaticRingPolymer2023, cookSpinMIntAlgorithmAccurate2026}
\begin{subequations} \label{conj-variables}
    \begin{align}
        &\frac{\textrm{d}}{\textrm{d} t} {\phi}_j = \frac{\partial H}{\partial r_{s,j}\cos\theta_j}  = H_z - H_x \frac{\cos \phi_j}{\tan \theta_j} - H_y \frac{\sin \phi_j}{\tan \theta_j}\text{,} \\
        &\frac{\textrm{d}}{\textrm{d} t} {r_{s,j}\cos\theta}_j = -\frac{\partial H}{\partial \phi_j} = \frac{1}{2} (H_xu_{y,j} - H_yu_{x,j})\text{,}
    \end{align}
\end{subequations}
which re-obtains Hamilton's EOM and is equivalent to the cross product in Eqn.~\eqref{u-heisenberg}.\cite{bossionNonadiabaticMappingDynamics2022} The electronic Liouvillian is 
\begin{align}
    \mathcal{L}_{\textrm{elec}} = \sum_l^N \frac{\textrm{d} \phi_l}{\textrm{d} t}  \frac{\partial}{\partial \phi_l} + \frac{\textrm{d} r_{s,l}\cos\theta_l}{\textrm{d} t} \frac{\partial}{\partial r_{s,l} \cos \theta_l} \textrm{,}
\end{align}
where we follow the convention set in the Matsubara paper and by Zwanzig,\cite{Hele2015, Zwanzig2001} and define the Liouvillian without the prefactor of the imaginary unit, $i$, such that it is real.
To prove conservation, one needs to show that the expectation value of $C_{AB}(t)$, where $A = \mathbb{I}$, is constant.\cite{Hele2015, Ananth2013, Hele2016, cookElectronicPathIntegral2025} This can be done by showing that the derivative is zero, noting that
\begin{align} \label{C_deriv_l_acting_b}
    \dot C^{[N]}_{\mathbb{I}B}(t) &= \frac{1}{Z} \int [e^{-\beta_N \bo{V}}]_{\bar{s}} \overrightarrow{ \mathcal{L}_\mathrm{elec}}  [\hat{B}(t)]_{s} \ \mathrm{d}\{\bo{U}_j\}  \text{,}
\end{align}
where $\mathcal{L}_\mathrm{elec}$ is the electronic Liouvillian. As it only consists of first-order derivatives,\cite{Hele2016} $\overrightarrow{\mathcal{L}_\mathrm{elec}} = - \overleftarrow{\mathcal{L}_\mathrm{elec}}$, such that the Liouvillian can act on the zero-time $[e^{-\beta_N \bo{V}}]_{\bar{s}}$ term. Therefore, we wish to show that 
\begin{align}
    \mathcal{L}_{\mathrm{elec}} [e^{-\beta_N \bo{V}}]_{\bar{s}} = 0 \textrm{.}
\end{align}

\subsubsection{Single Bead}
For a single bead, the SW transform of the single-bead Boltzmann distribution is 
\begin{align}
     [e^{-\beta_N \bo{V}}]_{\bar{s}} = \mathrm{Tr} \left[ e^{-\beta_N V_0} e^{-\beta_N \bo{H} \cdot \boldsymbol{\hat{\sigma}}/2} \hat{\omega}_{\bar{s}}\right] = \frac{1}{2} \mathrm{Tr} \left[ e^{-\beta_N \bo{V}} \left( \mathbb{I} + \bo{u} \cdot \boldsymbol{\hat{\sigma}} \right) \right] \textrm{.}
\end{align}
We define the derivatives of $\bo{u}$ with respect to the conjugate variables as\cite{cookSpinMIntAlgorithmAccurate2026}
\begin{subequations} \label{u_deriv_conj_var}
\begin{align}
    \bo{v}_1 &= \frac{\partial \bo{u}}{\partial r_s \cos\theta} = \begin{bmatrix}
        -2\cos\phi/\tan\theta \\
        -2\sin\phi/\tan\theta \\
        2
    \end{bmatrix} \text{,} \\
    \bo{v}_2 &= \frac{\partial \bo{u}}{\partial \phi} = \begin{bmatrix}
        -u_y\\
        u_x \\
        0
    \end{bmatrix} \text{,}
\end{align}
\end{subequations}
such that 
\begin{align} \label{sm-eta}
    \mathcal{L}_{\mathrm{elec}} [e^{-\beta_N \bo{V}}]_{\bar{s}} = (\bo{H}\times \bo{u}) \cdot \boldsymbol{\eta}^{\bar{s}} \textrm{,}
\end{align}
where $\boldsymbol{\eta}^{\bar{s}}_{i} = \mathrm{Tr}[e^{-\beta_N \bo{V}}\sigma_i]/2$. It should be noted that (as expected) acting the Liouvillian on $\bo{u}$ results in Heisenberg's EOM. $\bo{V}$ can be rewritten as
\begin{subequations}
\begin{align}
    \bo{V} &= \frac{(V_1 + V_2)}{2}\mathbb{I} + \frac{V_1 - V_2}{2}\hat{\sigma}_z + \Delta \hat{\sigma}_x + \Delta^* \hat{\sigma}_y \\
    &= \bo{V}_0 + \frac{1}{2}\sum_{i=1}^3 H_i \hat{\sigma}_i \textrm{,}
\end{align}
\end{subequations}
which is equivalent to Eqn.~\eqref{V_traceless} where $\bo{V}$ was decomposed into $\bo{V}_0$ (the identity-like term) and $\bar{\bo{V}}$ (the second term in the above) such that the exponential is 
\begin{subequations}
\begin{align}
    e^{-\beta_N \bo{V}} 
    &= e^{-\beta_N {\bo{V}_0}} \left[ \cosh \left( \beta_N |\bo{H}|/2\right)\mathbb{I} - \frac{\sinh \left( \beta_N |\bo{H}| /2 \right)}{|\bo{H}|} \bo{H} \cdot \boldsymbol{\hat{\sigma}}\right] \\
    &= a \mathbb{I} + b \bo{H} \cdot 
    \boldsymbol{\hat{\sigma}} \textrm{,}
\end{align}
\end{subequations}
which is obtained through Taylor expansion and, for simplicity, is rewritten in terms of constants $a =  e^{-\beta_N {\bo{V}_0}} \cosh \left( \beta_N |\bo{H}|/2\right)$ and $b = -e^{-\beta_N {\bo{V}_0}} \sinh \left( \beta_N |\bo{H}| /2 \right)/|\bo{H}|$. Hence 
\begin{subequations}
\begin{align}
    \boldsymbol{\eta}^{\bar{s}}_{i}  &= \frac{1}{2} \mathrm{Tr} \left[ a\hat{\sigma}_i + b \sum_{j=1}^3  H_j\hat{\sigma}_j\hat{\sigma}_i \right] \\
    &= \frac{1}{2}a\mathrm{Tr} \left[\hat{\sigma}_i \right] + \frac{1}{2} \sum_{j=1}^3 bH_i \mathrm{Tr}\left[ \hat{\sigma}_i\hat{\sigma}_j \right]  \\
    &= b H_i \textrm{,}
\end{align}
\end{subequations}
as $\mathrm{Tr}[\hat{\sigma}_i]=0$ and $\mathrm{Tr}[\hat{\sigma}_i\hat{\sigma}_j] = 2\delta_{ij}$. Overall 
\begin{subequations}
\begin{align}
    \mathcal{L}_{\mathrm{elec}} [e^{-\beta_N \bo{V}}]_{\bar{s}}  &=  b (\bo{H} \times \bo{u}) \cdot \bo{H} = 0 \textrm{,}
\end{align}
\end{subequations}
which is zero as a vector is orthogonal to its cross product. For a single bead, the distribution is conserved in $\bo{u}$.

\subsubsection{Multiple Beads}
\begin{figure}[b]
    \centering
    \includegraphics[width = 0.45\textwidth]{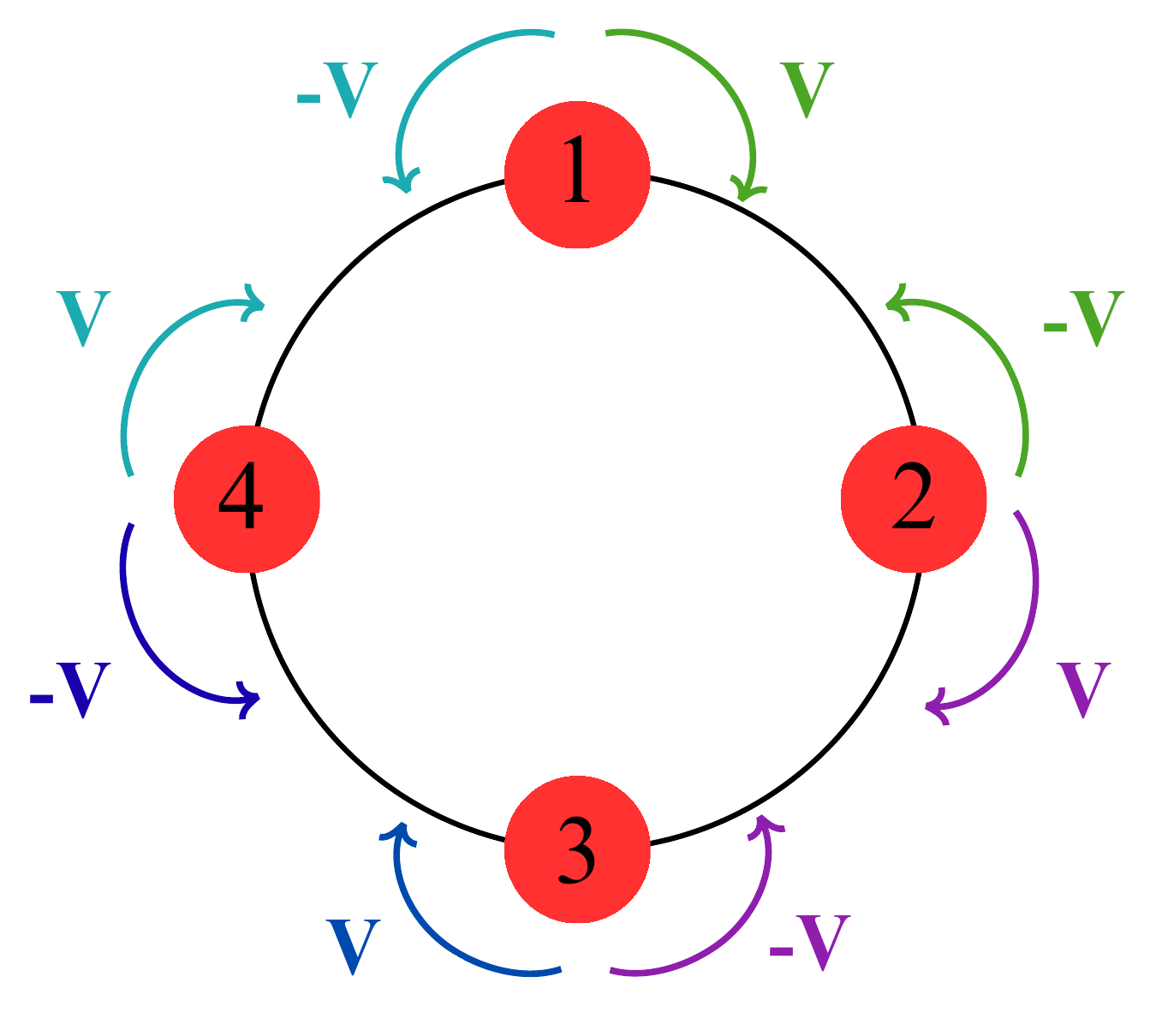}
    \vspace{-10pt}
    \caption{\label{fig:cancel} Schematic for acting the Liouvillian operator with 4 beads. The smooth lines are the imaginary-time evolution where the red circles indicate evaluation of $e^{\beta \bo{V}}\omega_{\bar{s},l}$ for the $l$-th bead. The two terms arising from green cyclic products will cancel and so on, until completing the path-integral loop results in full cancellation. }
\end{figure}

For multiple beads, the SW transform of the Boltzmann distribution is 
\begin{align}
    \mathcal{L}_{\mathrm{elec}} [e^{-\beta_N \bo{V}}]_{\bar{s}} = \sum_{l=1}^N (\bo{H}\times \bo{U}_l) \cdot \boldsymbol{\eta}^{\bar{s}}_{l} \textrm{,}
\end{align}
where 
\begin{align}
    \boldsymbol{\eta}^{\bar{s}}_{l} = \frac{1}{2} \mathrm{Tr}[e^{-\beta_N \bo{V}}\sigma_i\Gamma^{\bar{s}}_l] \textrm{,}
\end{align}
with $\Gamma^{\bar{s}}_l = \prod_{k = l+1}^{k= l-1} e^{-\beta_N\bo{V}}\hat{\omega}_{\bar{s},l} $ using cyclic notation as in Eqn.~\eqref{generalised-KT-cf}. For the $l$-th bead 
\begin{align} \label{single-bead-u}
    (\bo{H} \times \bo{U}_l) \cdot \boldsymbol{\eta}^{\bar{s}}_{l}  = \frac{1}{2}\mathrm{Tr}\left[ e^{-\beta_N \bo{V}} \left\{(\bo{H} \times \bo{U}_l) \cdot \hat{\boldsymbol{\sigma}}\right\} \Gamma^{\bar{s}}_{l} \right] \textrm{.} 
\end{align}
This can be written as a commutator due to
\begin{subequations}
    \begin{align}
        (\bo{H} \times \bo{U}_l) \cdot \hat{\boldsymbol{\sigma}} &= f_{ijk}H_i u_{j,l}\hat{\sigma}_k \textrm{,} \\
        [\bo{H}\cdot \hat{\boldsymbol{\sigma}}, \bo{U}_l\cdot \hat{\boldsymbol{\sigma}}] &= H_i u_{j,l} [\sigma_i, \sigma_j ] = 2i f_{ijk}H_i u_{j,l} \hat{\sigma}_k \textrm{,}
    \end{align}
\end{subequations}
where $f_{ijk}$ are the totally antisymmetric structure constants and Einstein summation is used. For the two-state case, these are equivalent to the Levi-Civita symbol which for cyclic permutations of $\{123\}$ is $+1$ and $-1$ otherwise. Hence
\begin{align} \label{commutator}
    (\bo{H} \times \bo{U}_l) \cdot \hat{\boldsymbol{\sigma}} = \frac{1}{2i}[\bo{H}\cdot \hat{\boldsymbol{\sigma}}, \bo{U}_l\cdot \hat{\boldsymbol{\sigma}}] \textrm{.}
\end{align}
We note a few interesting things, as $\bo{V} = \bo{V}_0 \mathbb{I} + (\bo{H}\cdot \hat{\boldsymbol{\sigma}})/2 $ then $(\bo{H}\cdot \hat{\boldsymbol{\sigma}})/2 = \bo{V}- \bo{V}_0 \mathbb{I}$ and one can reinsert the half-identity which does not change the commutator to obtain the kernel. Therefore
\begin{align}
    \frac{1}{2}(\bo{H}\times \bo{U}_l) \cdot \boldsymbol{\sigma} = -i \left[ \bo{V}, \frac{1}{2} \left( \bo{U}_l \cdot \hat{\boldsymbol{\sigma}} \right) \right] = -i [\bo{V}, \omega_{\bar{s}, l}] \textrm{.}
\end{align}
Inserting this into Eqn.~\eqref{single-bead-u} and expanding the commutator results in two terms per bead. A positive term where $\bo{V}$ appears to the right of $\omega_{\bar{s},l}$ and a negative term where it is to the left. The $l$-th bead positive term will therefore cancel with the negative term for the $(l+1)$-th bead as $\bo{V}$ commutes with $e^{-\beta\bo{V}}$. This is similar to the argument presented in Ref.~[\!\citenum{cookElectronicPathIntegral2025}], where acting the  Liouvillian operator `pulls' down a $\pm \bo{V}$ either side of each bead, and is outlined in Figure~\ref{fig:cancel}.
Overall 
\begin{align}
    \mathcal{L}_{\mathrm{elec}} [e^{-\beta_N \bo{V}}]_{\bar{s}} = \sum_{l=1}^N  -i\mathrm{Tr} \left[ e^{-\beta_N \bo{V}} [\bo{V}, \omega_{\bar{s}, l}]\Gamma^{\bar{s}}_l   \right]  =0 \textrm{,}
\end{align}
as summing over all beads results in all terms cancelling out.

When the CF is calculated, the imaginary-time part is calculated from the initial variables and the real-time from the time-evolved variables which can be calculated from all beads, all normal modes or truncated in the higher normal modes. Including all normal modes at finite time will also conserve the QBD as the back-transform can be utilised to obtain this bead representation.\cite{cookElectronicPathIntegral2025} One can additionally investigate the Liouvillian acting on the time-evolved part as in Eqn.~\eqref{C_deriv_l_acting_b}
\begin{align}
    \mathcal{L}_{\mathrm{elec}} [\hat{B}(t)]_{s}  &= \frac{1}{N} \mathcal{L}_{\mathrm{elec}}\left[ \sum_{j}^N B_{0} + \bo{B}\cdot \bo{U}_j(t) \right] \textrm{,}
\end{align}
as $\bo{B}$ is the same for all beads and is a linear operator. Acting the Liouvillian results in Heisenberg's EOM such that
\begin{subequations}
\begin{align}
     \mathcal{L}_{\mathrm{elec}} [\hat{B}(t)]_W
     &= \frac{1}{N} \mathcal{L}_{\mathrm{elec}} \left[ \sum_{j}^N B_{0} \right] + \mathcal{L}_{\mathrm{elec}}\left[\bo{B} \cdot \bar{\bo{U}}(t) \right] \\
     &=  \bo{B} \cdot \left[ \bo{H} \times \bar{\bo{U}}(t) \right] \textrm{,}
\end{align}
\end{subequations}
where as the normal mode propagation is separable, no higher normal modes contribute. As $[\hat{B}(t)]_W $ can be written as a function of only the lowest normal mode and the evolution of normal modes is separable, such that acting with the Liouvillian on the centroid results in propagation of only the centroid, truncating in the time-evolved normal modes will result in an exact correlation function which conserve the initial time distribution provided that the observable can be written as a function of only the lowest normal mode (which is the case for state occupancy and the difference between state occupancy as discussed above).

We have shown that acting with the Liouvillian on the initial-time distribution (using all beads) conserves the distribution. This is equivalent to exact real-time propagation, which by acting the Liouvillian on the real-time propagation we determine is only dependent on the centroid due to separable electronic normal mode evolution. Therefore, using all beads at initial-time and propagating only the centroid will result in exact, distribution conserving dynamics.

Hence, for an electronic-only system, we find that the spin-mapping representation algebraically satisfies all the desirable metric criteria which the MMST representation does not.\cite{cookElectronicPathIntegral2025}

\section{Results} \label{chap3-results}
Here, we computationally investigate the algebraic results. Specifically, we provide distributions of the spin-mapping normal modes which confirm the electronic constraint on the difference in the state populations. Following this, for an electronic-only system, we show that using only the spin-mapping centroid results in the exact first state population auto-CF and QBD conservation through comparison to the exact KT result. 

The potential energy matrix for a two-level electronic-only system is
\begin{align} \label{elec-only-pot}
    \bo{V} = \begin{bmatrix}
        \alpha & \Delta \\
        \Delta & -\alpha
    \end{bmatrix} \text{,}
\end{align}
where we have two sets of parameters to model a symmetric and asymmetric potential as tabulated in Table~\ref{param} and utilise atomic units where $\beta=1$. All calculations are for $N=8$ or $N=16$ beads, as detailed in figure captions. 
\begin{table} [H]
\caption{\label{param} Parameters for the symmetric and asymmetric potentials modelled. }
\begin{center}
    \begin{tabular}{c|c|c}
      \toprule 
      \ \textbf{Potential} \ & $\qquad \alpha \qquad $ & $\qquad  \Delta \qquad $ \\
      \midrule 
      \rowcolor[HTML]{C0C0C0} \ Symmetric \ & 0 & 1  \\
     \  Asymmetric \ & 1 &  1 \\
      \bottomrule 
    \end{tabular}
\end{center}
\end{table}
\vspace{-10pt}
We numerically test the distribution of the normal modes of $\bo{u}$ for narrowing which shows that the distribution constrains the normal modes of $\bo{u}_z$ which then impacts $\bo{u}_x$ and $\bo{u}_y$ as the elements of the spin-vector are not independent. We then investigate the first electronic state population auto-CF and the conservation of the QBD for $N$-beads and centroid-only propagation. The CFs are accurately captured using only the spin-mapping centroid whilst conserving the QBD for the spin-mapping representation.


\subsection{Normal Mode Distribution}

\begin{figure}[t]
    \centering
    \includegraphics[width = 0.85\textwidth]{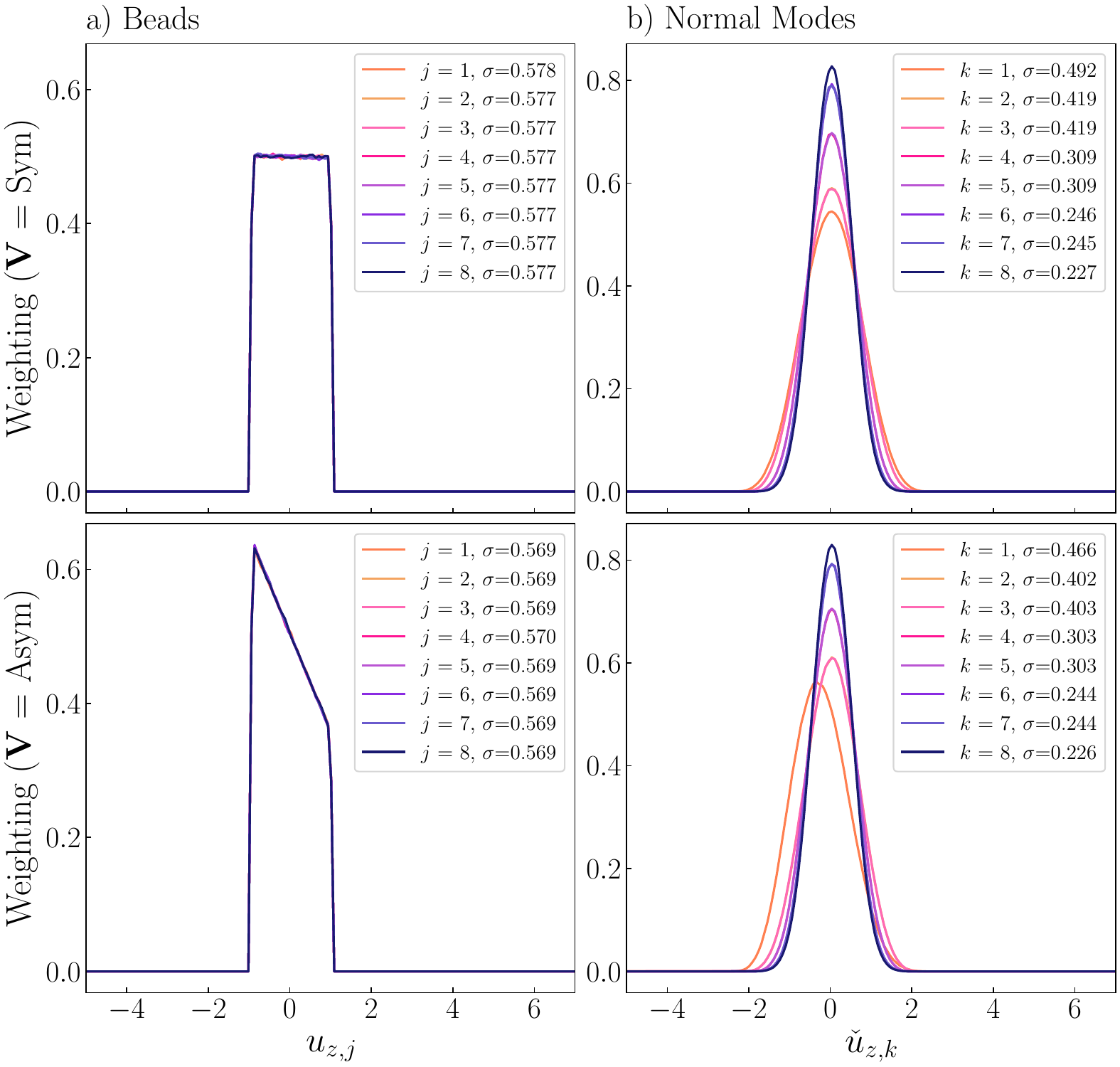}
    \vspace{-10pt}
    \caption{\label{fig:beads/modes_u_z_sm} Histogram of the distributions of a) beads and b) normal modes of $\bo{u}_z$ with the standard deviation in the legend for a symmetric (top row) and an asymmetric (bottom row) potential with spin-mapping sampling for 8 beads. We see that Boltzmann penalty acts as a constraint and the distribution of the higher normal modes narrows, whereas the distribution of beads does not.}
\end{figure}

The metric that shows narrowing in the higher normal modes is the difference between the state probabilities, $\bo{u}_z$, as seen in Figure~\ref{fig:beads/modes_u_z_sm} where $r_s= 1/2$ for 8 beads. The bead distributions on the left are the identical but the normal modes distributions on the right narrow with increasing normal mode index. The narrowing in the higher normal modes indicates the constraint arising from the penalty accrued when `hopping' between states. 

For the symmetric potential (top row), the initial sampling of beads is uniform between $-2r_s$ and $2r_s$. When modelling an asymmetric potential (bottom row), the sampling is biased to the lower energy state (when $\theta = \pi$) such that a linear decreasing ramp between $-2r_s$ and $2r_s$ is seen in the bead distributions.

As the constraint arises in the normal modes of $\bo{u}_z =2r_s\cos\theta$ and as $\bo{u}_x$ and $\bo{u}_y$ are functions of $\sin\theta$, the constraint also appears in the normal modes of $\bo{u}_x$ and $\bo{u}_y$ as seen in Figure~\ref{fig:modes_u_x_u_y} in Appendix~\ref{additional-distributions-sm}. However, the constraint discussed in Section~\ref{scaling-of-the-boltzmann-factor-with-N} relies on each bead being entirely in one state or the other, such that the off-diagonal elements of the kernel are zero. This is generally not the case as the spin-vector is in superposition of both states, and full characterisation of this constraint and how it impacts the other elements of the spin-vector is left as future work. Nevertheless, the results here suggest that it might be possible to integrate out the higher normal modes of the spin-mapping electronic dof to obtain a method which conserves the QBD, in the same way that the higher normal modes of the nuclear dof are integrated out in conventional Matsubara dynamics.\cite{Hele2015}

\subsection{Spin-Mapping Correlation Function}

\begin{figure}[b]
    \centering
    \includegraphics[width = 0.85\textwidth]{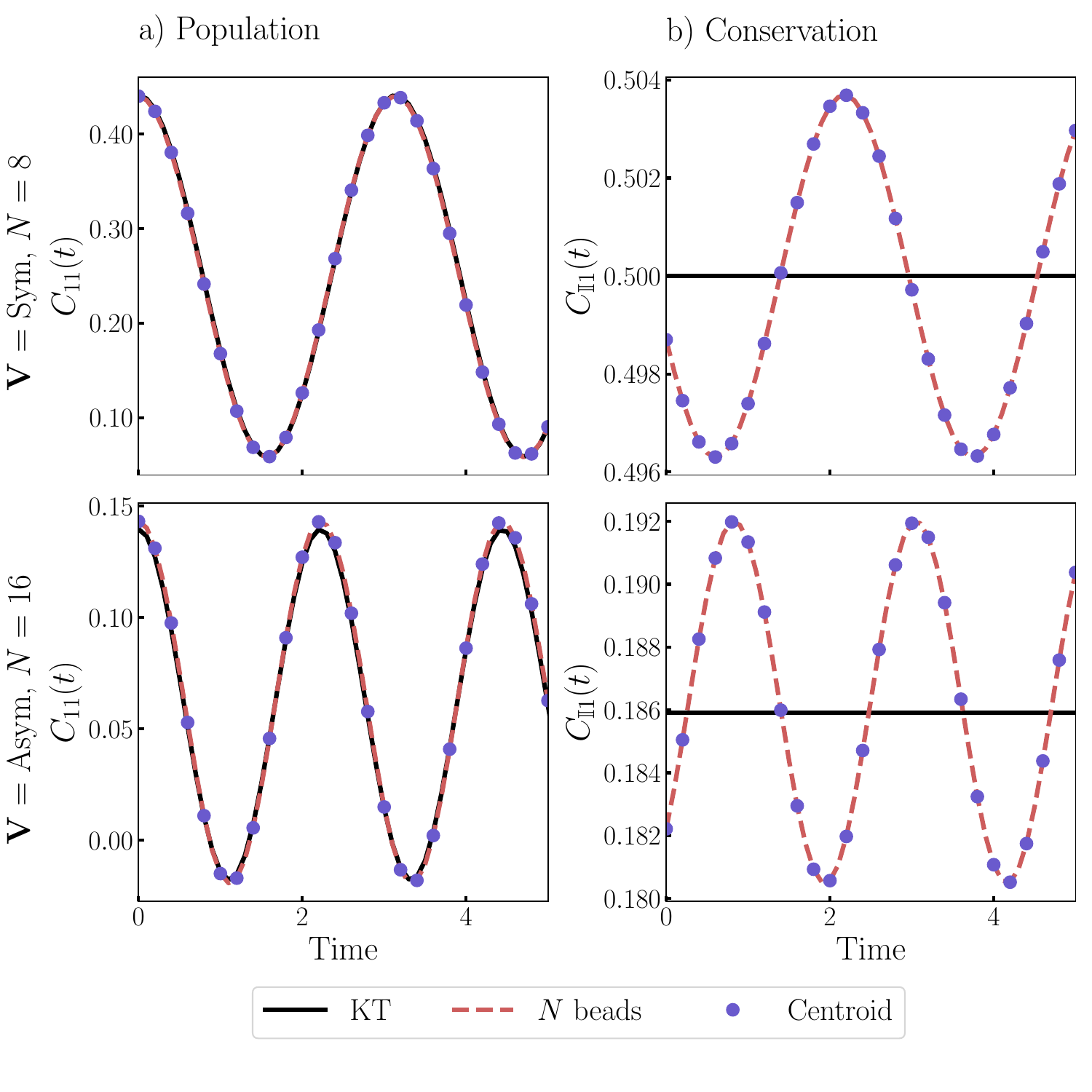}
    \vspace{-20pt}
    \caption{\label{fig:16-asym-spin-map} a) Autocorrelation function and b) conservation of the first state population with a symmetric (top row) and asymmetric (bottom row) potential for $N=8$ and $N=16$ beads respectively. The exact Kubo-transformed result (KT, black) is compared to a full $N$-bead calculation in spin-mapping beads (red dashed) and centroid-only (purple circle). All lines converge on the Kubo-transformed result with oscillations of 0.004 (symmetric) and 0.006 (asymmetric) in the conservation plot.}
\end{figure}

The spin-mapping auto-CFs is shown on the left hand side of Figure~\ref{fig:16-asym-spin-map}, for a symmetric and asymmetric potential. The symmetric calculation uses 8 beads with 4 million trajectories, such that direct comparison with Ref.~[\!\citenum{cookElectronicPathIntegral2025}] is possible, and the asymmetric calculation uses 16 beads with 8 million trajectories. It can be seen that including all beads and only propagating the centroid, $\bar{\bo{u}}$, results in identical correlation functions compared to the exact KT result, calculated using the trapezium rule for $N$ beads. 

\begin{figure}[b]
    \centering
    \includegraphics[width = 0.6\textwidth]{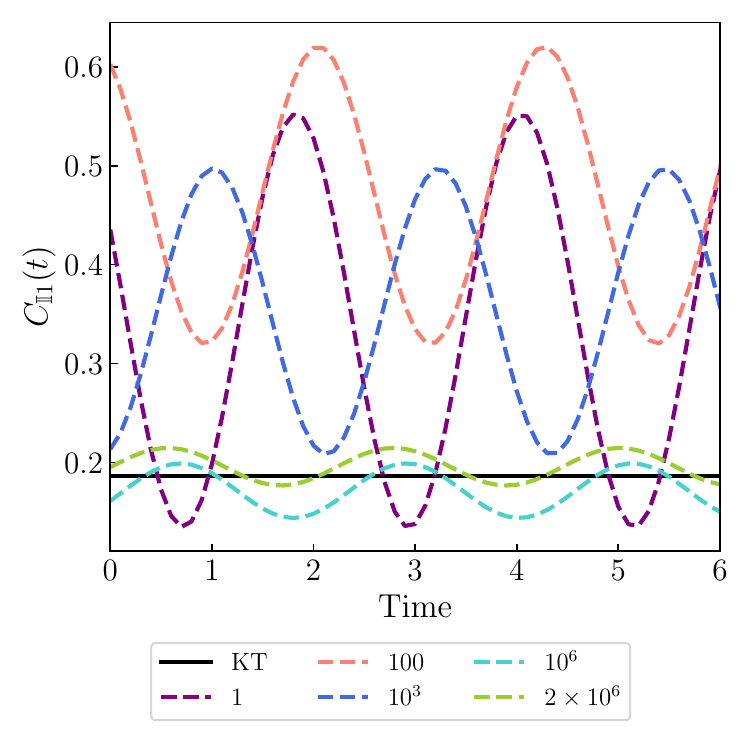}
    \vspace{-20pt}
    \caption{Correlation function for the conservation of electronic population of the first state (asymmetric potential, 16 beads) with the exact Kubo-transformed result (KT, black) compared to the centroid-only calculation with different numbers of trajectories; 1 (purple), 100 (orange), $10^3$ (blue), $10^6$ (light blue), $2 \times 10^6$ (green). Increasing the number of trajectories improves the accuracy of the correlation function and the amplitude of oscillations decreases, showing that the deviation from the KT result is statistical.}
    \label{fig:16-spin-map-converge}
\end{figure}

The corresponding conservation of the distribution (right) shows that the centroid-only result is identical to including all beads and has small oscillations around the exact KT results (note the scale on the $y$ axis). By comparing the symmetric and asymmetric results, which are for 8 and 16 beads respectively, we see that including more beads slightly increases the size of these oscillations due to more trajectories being required to converge the results. In Figure~\ref{fig:16-spin-map-converge}, we show that including more trajectories improves the conservation of the distribution (asymmetric, 16 bead calculation) by reducing the amplitude of oscillations for the centroid-only result. These errors are therefore statistical as including more trajectories converges the result to the exact KT result. Hence, propagating only the centroid results in accurate dynamics and conservation of the QBD.

An alternative approach, and one that is presented in the Supplementary Material, is to sample MMST variables and compute the MMST CF from propagation of the spin-mapping variables, utilising the transformation between the representations in Eqn.~\eqref{mmst-spin}. The spin-vector is propagated as in Eqn.~\eqref{u-prop} and the time-evolved observable also becomes dependent only on the spin-mapping centroid. The spin-mapping normal modes calculated from the sampled MMST variables are also constrained. We stress that this is not the same as the results in Ref.~[\!\citenum{cookElectronicPathIntegral2025}] as the MMST representation is used for sampling but the dynamics and CF are computed from transformation into spin-mapping variables and spin-mapping normal modes.  

However, there are a few advantages to the spin-mapping CF over transforming the MMST CF into spin-mapping variables. Firstly, sampling a sphere is more efficient than Gaussian MMST distributions. This is highlighted by Fig.~S.2 in the Supplementary Material for MMST sampling (asymmetric, 8 beads, 96 million trajectories), which can be compared to the spin-mapping results (asymmetric, 16 beads, 8 million trajectories) in Figure~\ref{fig:16-asym-spin-map} which has tighter convergence with significantly fewer trajectories for more beads. Hence, the spin-mapping sampling allows for much larger systems to be probed by reducing the amount of trajectories required for convergence. Other advantages of spin-mapping include limiting ZPE-leakage and invariance to the choice of splitting of the potential which has been shown to improve the accuracy of results.\cite{Bossion2021, Runeson2019}

\section{Conclusions} \label{conclusions}
In this work, we found that the electronic metric constrained by the QBD is the difference in state populations, which maps to the $z$-component of the spin-vector for a two-state system. For the electronic-only system, the observables simplify into a function of only the centroid of the spin-vector. Algebraically and numerically, we have shown propagating only the centroid of the spin-vector results in conservation of the distribution. These properties were not obtained for the MMST normal modes in Ref.~[\!\citenum{cookElectronicPathIntegral2025}]. Whilst one can convert the \textit{dynamics} of the previous MMST approach into spin-mapping and obtain exact results using only spin-mapping centroid, the spin-mapping CF is advantageous due to improved convergence when sampling a sphere. In Table~\ref{C3-table1}, the findings in relation to the criteria for a desirable electronic state metric in comparison to nuclear Matsubara modes and the previously investigated MMST electronic normal modes are tabulated.\cite{Hele2015,cookElectronicPathIntegral2025} 

\begin{table} [h]
\caption{\label{C3-table1}Summary of the results in relation to the criteria for a desirable electronic normal modes metric in comparison to nuclear Matsubara modes. The middle column summarises the results of the MMST electronic normal modes investigated in Ref.~[\!\citenum{cookElectronicPathIntegral2025}]. The right column is the central results of this article investigating the spin-mapping electronic normal modes.}
\begin{center}
    \begin{tabular}{p{0.435\linewidth}|c|c|c}
      \toprule 
      \textbf{Criterion} & \makecell[c] {\textbf{Nuclear} \textbf{Matsubara} \\ {\textbf{Modes}}} & \makecell[c] {\textbf{MMST}  \textbf{Electronic } \\ {\textbf{Modes}}}  & \makecell[c] {\textbf{SM}  \textbf{Electronic} \\ {\textbf{Modes}}}\\
      \midrule 
      \rowcolor[HTML]{C0C0C0} Observable function of finite number of NM & \cmark & \xmark & \cmark \\
      Constrained higher NM & \cmark & \xmark & \cmark  \\
      \rowcolor[HTML]{C0C0C0} Truncated single trajectory QBD conservation & \cmark & \xmark & \cmark \\
      Truncated ensemble QBD conservation & \cmark & \cmark (numerically) & \cmark \\
      \rowcolor[HTML]{C0C0C0} Accurate dynamics upon truncation& \cmark & \xmark  & \cmark\\
      \bottomrule 
    \end{tabular}
\end{center}
\end{table}

It is clear that the properties of the electronic spin-mapping normal modes are promising for finding a nonadiabatic Matsubara dynamics that conserves the QBD. It is unlikely that when reintroducing nuclear dof, the electronic normal modes will remain uncoupled and observables will simplify to be functions of only the centroid. However, as the higher normal modes of the spin-vector appear to be constrained by the QBD, it may be possible to integrate them out (as is done in single-surface Matsubara dynamics).\cite{Hele2015} For this, full characterisation on how this constraint affects the normal modes of the $x$ and $y$ components of the spin-vector and the behaviour for larger numbers of beads is required. Following this, extension to a general number of electronic states using the $SU(F)$ Lie group would be desirable.\cite{Runeson2020, bossionNonadiabaticMappingDynamics2022, bossionNonadiabaticRingPolymer2023} 

These results provide reasoning for the recent success of spin-mapping compared to MMST, and while they do not yet comprise of a generally-applicable method, this work should be utilised to significantly aid the future derivation of highly accurate, QBD-conserving nonadiabatic dynamics methods.

\section*{Supplementary Material}

\section*{Acknowledgements}
TJHH acknowledges a Royal Society University Research Fellowship URF\textbackslash R1\textbackslash 201502 and renewal URF\textbackslash R\textbackslash 251018. LEC acknowledges a University College London studentship. JRR acknowledges funding from the Engineering and Physical Sciences Research Council [grant number EP/Z534882/1]. The authors thank Stuart Althorpe for helpful discussions.

\section*{Author Declaration}
The authors have no conflicts to disclose.

\section*{Data Availability} \label{data-avaliability}
The data that support the findings of this study are openly available in UCL Research Data Repository at http://doi.org/[to be inserted], reference number [to be inserted].

\appendix
    \renewcommand{\thesubsection}{\Alph{subsection}}
    \section*{Appendices}
    \addcontentsline{toc}{section}{Appendices}
    \numberwithin{equation}{subsection}

\subsection{Spin-Mapping Sampling and Partition Function} \label{sampling-elec}
The partition function and sampling used here is very similar to Ref.~[\!\citenum{bossionNonadiabaticRingPolymer2023}] without any nuclear dof. The partition function in Eqn.~\eqref{partiton} under the SW transform becomes
\begin{align}
    Z = \int \mathrm{Tr}\left[\prod_j^N e^{-\beta_N H}\omega_{s,j}\right] \ \mathrm{d}\{\boldsymbol{u}_j\} \textrm{,}
\end{align}
where $\mathrm{d}\{\boldsymbol{u}_j\} = \frac{1}{(2\pi)^N} \prod_j^N \sin \theta_j \ \mathrm{d}\theta_j  \ \mathrm{d}\phi_j$ and defining a distribution
\begin{align}
    \rho = \frac{1}{(2\pi)^N}|Q| \prod_j^N \sin \theta_j \textrm{,}
\end{align}
where $Q= \mathrm{Tr}[\prod_j^N e^{-\beta_N H}\omega_{s,j}]$ is the complex weighting such that
\begin{align}
    Z = \left\langle  \frac{Q}{|Q|} \right\rangle_{\rho} \textrm{,}
\end{align}
The distribution is sampled using a Metropolis-Hastings algorithm, sampling $\phi_j \in \{0,2\pi\}$ and $\theta_j = \arccos(z_j)$ for $z_j \in \{-1,1\}$ and weighting by $|Q|$. 

Therefore, the correlation function is calculated as
\begin{subequations}
\begin{align}
    C_{AB}^{[N]} (t) &= \frac{1}{Z} \int \rho \frac{[e^{-\beta_N H} \hat{A}]_{\bar{s}}}{|Q|} \times [\hat{B}(t)]_{s} \mathrm{d}\{\boldsymbol{u}_j\} \\
    & = \frac{ \left\langle \frac{[e^{-\beta_N H} \hat{A}]_{\bar{s}}}{|Q|} \times [\hat{B}(t)]_{s} \right \rangle_{\rho}}{\left \langle  \frac{Q}{|Q|}  \right \rangle_{\rho}} \textrm{,}
\end{align}
\end{subequations}
by summing over many trajectories. 

\subsection{Additional Distributions} \label{additional-distributions-sm}

\begin{figure}[b]
    \centering
    \includegraphics[width =0.85 \textwidth]{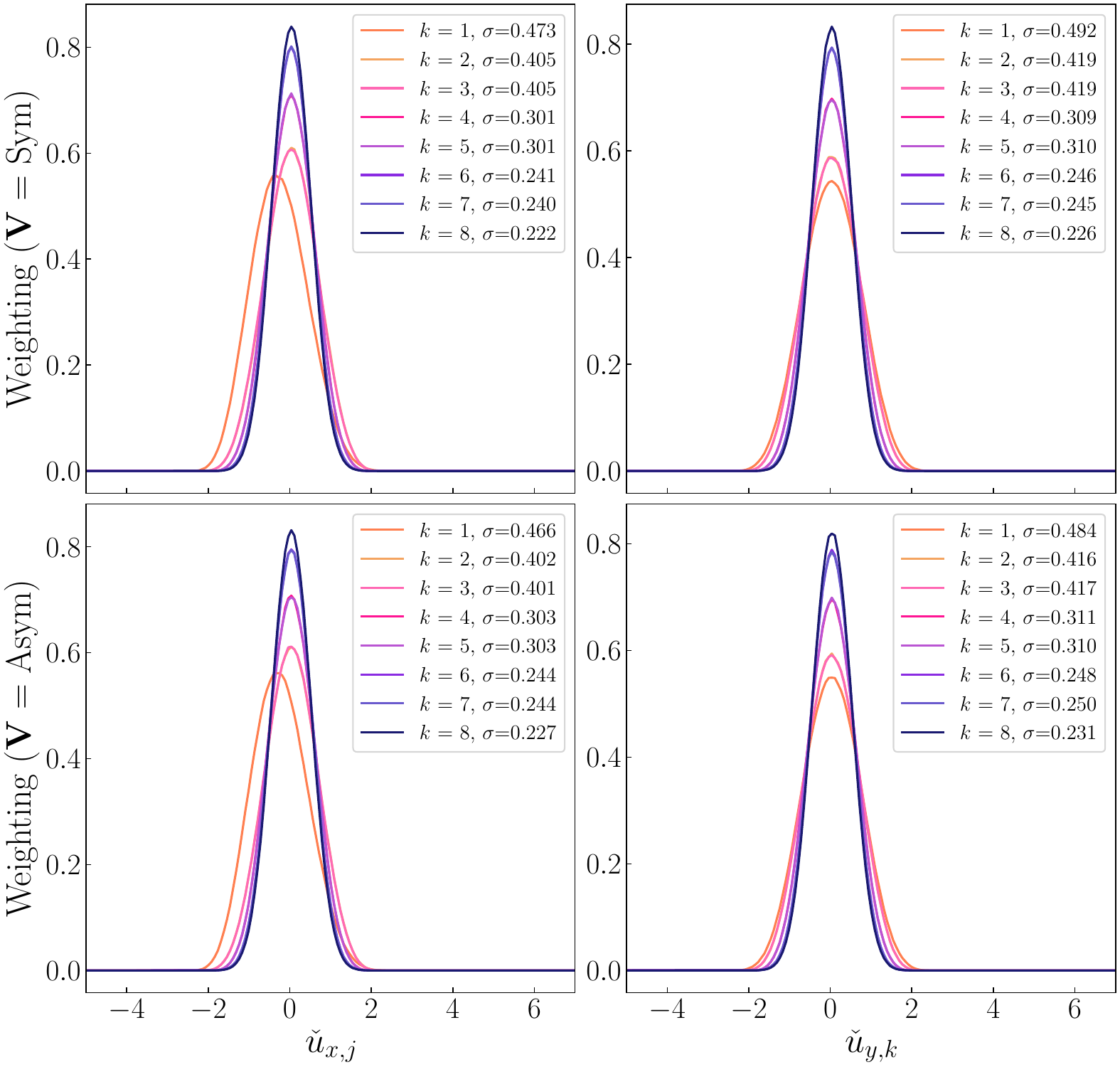}
    \vspace{-10pt}
    \caption{\label{fig:modes_u_x_u_y}Histogram of the distributions of normal modes of $\bo{u}_x$ (left) and $\bo{u}_y$ (right) with the standard deviation in the legend for a symmetric (top row) and asymmetric potential (bottom row) with spin-mapping sampling for 8 beads. We see in both plots that these distributions appear constrained (due to the narrowing), likely arising from the relationship with $\bo{u}_z$.}
\end{figure}

In Figure~\ref{fig:modes_u_x_u_y}, the distributions of the normal modes of $\bo{u}_x$ (left) and $\bo{u}_y$ (right) are presented for both a symmetric and asymmetric potential. It can be seen that the higher normal modes of both narrow and elongate suggesting that the constraint on $\bo{u}_z$ is passed on to these variables. 

It is interesting to note that the symmetric and asymmetric potentials can be decomposed in terms of Pauli spin-matrices 
\begin{align}
    \label{pot-spin-mat}
    \bo{V}_{\textrm{sym}} &= \hat{\sigma}_x  \\
    \bo{V}_{\textrm{asym}} &= \hat{\sigma}_x + \hat{\sigma}_z .
\end{align}
As the contributions of the $x$ and $y$ spin-matrices are the same for both potentials, there is no difference in the normal mode distributions when changing the potential. The co-efficient for $\hat{\sigma}_x$ is the same as for $\hat{\sigma}_z$ in the asymmetrical potential, which is why the $\check{u}_x$ normal mode distributions look the same as for the asymmetric $\check{u}_z$ in Figure~\ref{fig:beads/modes_u_z_sm}. Full algebraic characterisation of how the constraint affects these components of the spin-vector is left as future work.

\bibliography{rp2}

\end{document}